\documentclass[twocolumn,showpacs,preprintnumbers,aps,amsmath,amssymb]{revtex4}

\usepackage{graphicx}
\usepackage{dcolumn}
\usepackage{bm}
\usepackage{color}
\usepackage[left]{lineno}

\begin{document}

\preprint{}

\title{Modeling of Spin Metal-Oxide-Semiconductor Field-Effect-Transistor: \\
A Non-Equilibrium Green's Function Approach with Spin Relaxation}

\author{Tony Low$^{1}$, Mark S. Lundstrom$^{1}$ and Dmitri E. Nikonov$^{2}$}
\affiliation{
$^{1}$Department of Electrical and Computer Engineering, Purdue University, West Lafayette, Indiana 47906, USA \\
$^{2}$Technology Strategy, Technology and Manufacturing Group, Intel Corporation, 2200 Mission College Blvd., Santa Clara, California 95052, USA \\
}

\date{\today}

\begin{abstract}\textcolor{black}{
A spin metal-oxide-semiconductor field-effect-transistor (spin MOSFET), which combines a Schottky-barrier MOSFET with ferromagnetic source and drain 
contacts, is a promising device for spintronic logic. Previous simulation studies predict that this device should display a very high magnetoresistance (MR) ratio (between the cases of parallel and anti-parallel magnetizations) for the case of half-metal ferromagnets (HMF). We use the non-equilibrium Green's function (NEGF) formalism to describe tunneling and carrier transport in this device and to incorporate spin relaxation at the HMF-semiconductor interfaces. Spin relaxation at interfaces results in non-ideal spin injection. Minority spin currents arise and dominate the leakage current for anti-parallel magnetizations. This reduces the MR ratio and sets a practical limit for spin MOSFET performance. We found that MR saturates at a lower value for smaller source-to-drain bias. In addition, spin relaxation at the detector side is found to be more detrimental to MR than that at the injector side, for drain bias less than the energy difference of the minority spin edge and the Fermi level.} 
\end{abstract}

\pacs{72.25.2b, 73.23.2b, 73.40.Sx, 85.75.2d}

\maketitle

\section{\label{sec:level1}INTRODUCTION}

In recent years, a vigorous research effort to demonstrate spintronic devices \cite{wolf01,zutic04} has been pursued. One of the motivations has been that spin-based devices are identified as one of the most promising alternatives to traditional, charge-based logic devices by the International Technology Roadmap for Semiconductors \cite{itrs07}. Simulations have predicted that spintronic logic can scale in its size with smaller switching energy and less overall power dissipation than electronic logic \cite{nikonov08}.

The concept and operating principles of the first magnetic three-terminal device, i.e. the spin current modulator, was proposed by Datta and Das \cite{datta90} in 1990. It comprises of a gate controlling the spin precession in a semiconductor channel, a ferromagnetic (FM) source injecting highly polarized spins, and a FM drain detecting the spin polarization. The current depends both on the relative directions of magnetization of the source and the drain and on the gate bias. Without the gate field, the spin current modulator can acts as a giant magnetoresistance device \cite{fert94}. 
The gate exerts an effective magnetic field (Rashba field) \cite{bychkov84,meier07} which causes precession of the spins as they move along the channel. With enough channel length and strong enough Rashba field, the angle of spin precession can be varied from $0$ or $\pi$, generating the device ON and OFF states, respectively, for parallel magnetizations. Since the device operation involves a precise phase of the spins, it is desirable to suppress any scattering mechanisms, i.e., the device works best at low temperature and in samples with few defects. Physical realization of the Datta-Das spin current modulator is mainly impeded by the substantial difficulties of obtaining efficient room-temperature spin injection from the FM contacts into the semiconductors like GaAs \cite{adelmann05} and Si \cite{jonker07}. A recent realization of spin field effect transistor (FET) \cite{huang07} using hot-electron transport through FM thin films for all-electrical spin polarized injection and detection \cite{appelbaum07} with a Si channel is encouraging. There the electric field controls the transit time of electrons and thus its precession in the magnetic field. However it seems unlikely that the Datta-Das spin current modulator or any other spin precession devices can provide the ON/OFF current ratio comparable to traditional electronic metal-oxide-semiconductor field-effect-transistors (MOSFET).

Another type of a spin transistor, a spin MOSFET, was proposed by Sugahara and Tanaka \cite{sugahara04}. It is in essence a Schottky barrier (SB) MOSFET, where the source and drain are ferromagnetic. A half-metal ferromagnet (HMF) was employed in the original proposal, i.e., a material
having 100\% of electrons with one direction of spin at its Fermi level \cite{groot83}. This property is conducive to higher spin polarization of injected carriers. Silicon's mature technology base makes it a preferred choice for a channel material. Furthermore, the low spin relaxation rate due to its relatively small spin-orbit effects and negligible hyperfine interaction gives propagating electrons in Si a substantially long spin lifetimes \cite{tyryshkin05}. The authors of \cite{sugahara04} argued that spin MOSFETs might be used for high-density non-volatile memory, whose cell contains a single spin transistor, as well as for non-volatile, reconfigurable logic circuits \cite{tanaka07}. The difference between the spin MOSFET and a spin current modulator is that the former does not rely on the phase of the spin precession. Instead, the current is controlled by the height of the Schottky barriers, which is different for electrons with different spins. This is due to the fact that the states of electrons with spin along and opposite to the magnetization are split by the value of the exchange interaction in the ferromagnet. The role of the gate is to change the electrical potential and thus the thickness of the Schottky barriers. Together the directions of magnetization and the gate bias determine the current through the transistor.

In all types of spin transistors, the key device metric is the magnetoresistance (MR) ratio, i.e., the ratio of currents for parallel and anti-parallel magnetizations. It is a measure of the control of carrier transport by the magnetic state of the device. Theoretically, HMFs will have $100\%$ spin polarization (all the spins being aligned with the magnetization) as first predicted by density functional theory for NiMnSb \cite{groot83} and also supported by experiments such as spin-resolved positron annihilation \cite{hanssen90} and infrared reflectance spectroscopy \cite{mancoff99}. Therefore, one expects injection of  100\% spin polarized carriers from a HMF source, which would lead to an extremely large MR ratio in the spin MOSFET \cite{sugahara04}. \textcolor{black}{The class of Heusler alloys of type X-MnSb are excellent candidates for HFM \cite{galanakis02}. Genuine half-metallic interfaces of NiMnSb with III-V semiconductors (e.g. InP and CdS) were predicted in the $[111]$ direction with both HMF and semiconductor being anion terminated at the interface \cite{wijs01}. Alternatively, one could also consider non-magnetic semiconducting Heusler alloys (e.g. NiMbSb, NiScSb, NiTiSn and CoTiSb) which provide a smaller lattice mismatch with Heusler type HMF. Of particular importance is CoTiSb, a semiconductor with indirect bandgap with conduction energy minimum along the six-fold degenerate $\Gamma X$ symmetry lines, just like Si \cite{attema06}.}

However, high spin polarization of injected carriers has not yet been achieved with HMF. Experiments on spin-polarized photoemission \cite{bona85} and spin-polarized tunneling \cite{tanaka99} with HMF result in spin polarization values far below $100\%$. This is attributed to the presence of a `magnetically dead layer', i.e., an area close to the surface of the FM which is not ferromagnetic. Its random magnetization interacts with the spins of injected carriers and thus decreases their spin polarization. In particular, a recent first-principle study of half-metallic NiMnSb/CdS interface reveals that the NiMnSb surfaces are not half-metallic, even if they are stoichiometric and perfectly ordered \cite{wijs01}. Therefore, in order to estimate a realistic performance of spin MOSFET, one has to take into account spin relaxation processes at these interfaces. 

Many publications have addressed the simulation of spin-dependent transport. The drift-diffusion approach to spin transport with spin flip processes is reviewed in \cite{zutic06}. However, quantum tunneling processes cannot be handled by this method. The quantum conductance treatment had been applied \cite{bauer01} to simulation of a spin-flip transistor. Non-equilibrium Green's function (NEGF) treatment of tunneling in ferromagnetic metal-oxide multilayers with spin relaxation was described in \cite{yanik07,salahuddin07}. NEGF has also been applied to spin transport in carbon nanotubes \cite{wang03} and molecures \cite{rocha05}. \textcolor{black}{The contribution of our work is to treat quantum transport through metal and semiconductor structures with spin relaxation at the interfaces and under the influence of a self-consistent electrostatic field.} 

In this paper, we describe a full quantum-mechanical model for simulating carrier transport in a spin MOSFET, based on the Keldysh non-equilibrium Green's function approach \cite{datta97,haug96,mahan90}. In our model, we capture the physics of carrier injection and extraction, tunneling through Schottky barriers, quantum interference of electron wave reflections, and spin relaxation. The influence of the `magnetically dead layer' is incorporated via a scattering self-energy of interaction of spin of carriers and localized electrons, derived within the self-consistent Born approximation \cite{datta04,yanik07}. We quantify the effects of spin relaxation on the MR ratio of the spin MOSFET. 
The rest of this paper is organized as follows. In section II, we introduce the NEGF formalism with a detailed mathematical description of the physical quantities used in the formalism. In section III, we apply this approach to the study of spin MOSFET in the coherent regime (without spin relaxation). The bias dependence of MR ratio is explained in this section. In section IV, we examine the effect of the `magnetically dead layer' on spin transport in spin MOSFET. In section V, we explore the dependence of the MR ratio on the strength of interaction with a 'magnetically dead layer'. Conclusions are drawn in Section VI. In Appendix A, we present a derivation of the scattering self energy within the self-consistent Born approximation. In Appendix B, a simple ohmic model analysis of spin MOSFET with HMF and FM source/drain contacts is discussed. 

\section{\label{sec:level2}MODEL DESCRIPTION}

\begin{figure}[t]
\centering
\scalebox{0.35}[0.35]{\includegraphics*[viewport=45 120 780 595]{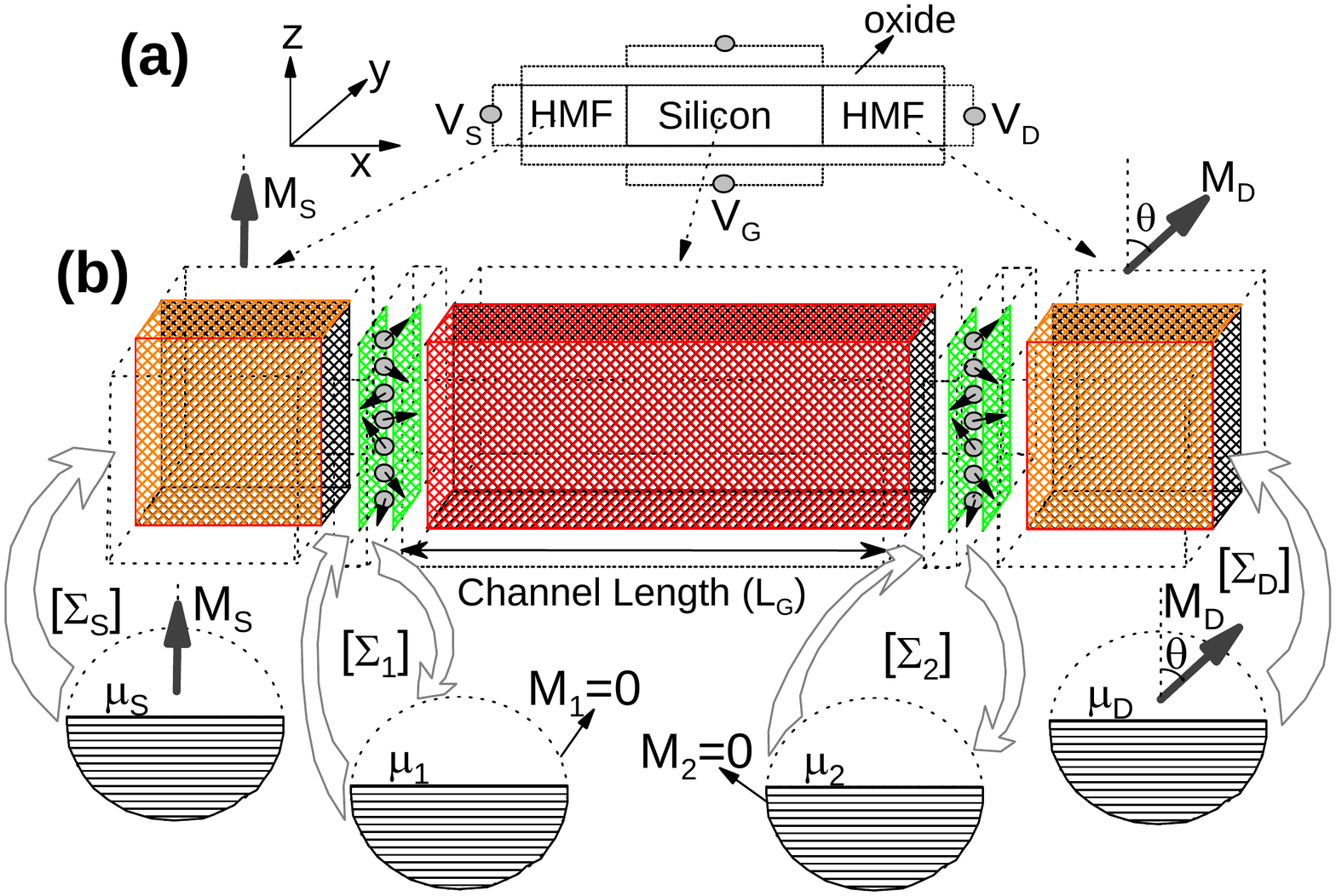}}
\caption{\footnotesize (a) A schematic illustration of the three terminal (i.e. gate contact $V_{G}$, drain contact $V_{D}$ and source contact $V_{S}$) spin-MOSFET device. A double-gate structure is employed. (b) A zoom-in illustration of the supercell used for the construction of device Hamiltonian depicting the various self-energies used in the calculation.}
\label{deviceschematic}
\end{figure}

A schematic drawing of a spin MOSFET is illustrated in Fig.\ref{deviceschematic}. \textcolor{black}{We employed here a double-gate structure with a thin film semiconductor for which the gate control of electrostatics is optimal. The channel is a semiconductor and the source/drain contacts are half-metal ferromagnets (HMF) \cite{attema06} with magnetization of $\mathbf{M}_{S}$ and $\mathbf{M}_{D}$, respectively. The device is large in the $y$ direction.} Therefore Fig.\ref{deviceschematic} depicts the supercell for our transport problem, where the wavefunction solution repeats periodically in the $y$ direction. We employ the effective mass approach to the description of the bandstructure. The Hamiltonian and the Schr$\ddot{o}$dinger equation reduce to the following Sturm-Liouville problem in the longitudinal plane ($\mathbf{r}=(x,z)$) (omitting the spin relaxation terms here);
\small
\begin{eqnarray}
\nonumber
H\Psi_{\sigma}(\mathbf{r})&\equiv& -\frac{\hbar^{2}}{2}\nabla_{r}\cdot\left[M^{-1}(\mathbf{r})\nabla_{r}\Psi(\mathbf{r})\right]\\
&&+\left[V(\mathbf{r})+\epsilon_{y}+\mathbf{M}(\mathbf{r})\cdot\mathbf{\sigma}\right]\Psi_{\sigma}(\mathbf{r})= \epsilon \Psi_{\sigma}(\mathbf{r})
\label{sys}
\end{eqnarray}
\normalsize   
where $M^{-1}(\mathbf{r})$ is the effective mass tensor, with diagonal elements $m_{x}^{-1}$ and $m_{z}^{-1}$ \footnote{The matrix elements of $M^{-1}(\mathbf{r})$ are $[M^{-1}]_{11}$$=$$m_{x}^{-1}(\mathbf{r})$, $[M^{-1}]_{12}$$=$$m_{xz}^{-1}(\mathbf{r})$, $[M^{-1}]_{21}$$=$$m_{zx}^{-1}(\mathbf{r})$ and $[M^{-1}]_{22}$ $=$ $m_{z}^{-1}(\mathbf{r})$. In this work, we set $m_{xz}(\mathbf{r})$=$m_{zx}(\mathbf{r})$=$0$}. We assume that the transport mass ($m_{x}$), transverse mass ($m_{y}$) and quantization mass ($m_{z}$) are spatially uniform within each material. $\mathbf{M}(\mathbf{r})$ is the magnetization at $\mathbf{r}$, $\mathbf{\sigma}=(\sigma_{x},\sigma_{y},\sigma_{z})$ are the Pauli spin matrices and $\nabla_{r}\equiv \partial_{x}\mathbf{i}+\partial_{z}\mathbf{k}$. 
The eigenvalue of $H$ is the total energy $\epsilon$, while $\epsilon_{y}=\hbar^{2}k_{y}^{2}/2m_{y}$ is the transverse energy corresponding to the transverse mode solution $\phi(y)=exp(ik_{y}y)/\sqrt{w}$. 
In this work, we use the finite difference approach to express our physical quantities in matrix representation \cite{datta97}. Assuming that the electrostatic potential is separable, i.e., can be represented as $V(\mathbf{r})\approx V_{1}(x)+V_{2}(z)$, we can further reduce our system Eq.~(\ref{sys}) to a one-dimensional problem \footnote{It is also theoretically possible to further decouple the Hamiltonian $H$ into two 1D problem when $V(\mathbf{r})$ can be written as $V(\mathbf{r})=V_{1}(x)+V_{2}(z)$ \cite{venugopal02}, commonly referred to as the 'mode space approach'. When the film thickness ($T_{b}$) is sufficiently thin, i.e. $<$$3nm$ for Si, the mode space approach is relatively accurate.}.

The Keldysh non-equilibrium Green's function (NEGF) approach \cite{datta97,haug96,mahan90} is a powerful technique in solving electronic transport in nanostructures having open boundary conditions. The infinite problem domain ($\Omega$) which consists of the semiconductor channel and two semi-infinite HMF leads is partitioned into an exterior ($\Omega_{e}$) and an interior domain ($\Omega_{i}$), where only the solutions within $\Omega_{i}$ is to be sought. In this work, $\Omega_{i}$ is composed of the semiconductor channel and a small segment of the half-metal ferromagnet at the source/drain. In the absence of scattering, the Green's function $G$ in $\Omega_{i}$ is written as,
\small
\begin{eqnarray}
G(\epsilon_{x})=\left[(\epsilon_{x}) I-H-V_{1}-\Sigma_{c}(\epsilon_{x})\right]^{-1}
\label{gfunction}
\end{eqnarray}
\normalsize   
where $\Sigma_{c}=\Sigma_{L}+\Sigma_{R}$ is the contacts self-energy \cite{datta97} and $\epsilon_{x}=\epsilon-\epsilon_{y}-\epsilon_{j}$ with $\epsilon_{j}$ being the subband energy of the $j$-th mode due to the $z$ confinement. For a one-dimensional lattice with a nearest neighbor coupling energy of $t$ and lattice spacing of $a$, the contact self-energy is written as,
\small
\begin{eqnarray}
\left[
\begin{array}{cc}
c_{i} & s_{i}\\
-s_{i} & c_{i}\\
\end{array}
\right]\left[
\begin{array}{cc}
-te^{ik_{i}^{\uparrow}a} & 0\\
0 & -te^{ik_{i}^{\downarrow}a}\\
\end{array}
\right]\left[
\begin{array}{cc}
c_{i} & s_{i}\\
-s_{i} & c_{i}\\
\end{array}
\right]^{\dagger}
\end{eqnarray}
\normalsize   
where $c_{i}=cos(\theta_{i}/2)$, $s_{i}=sin(\theta_{i}/2)$, $i$ labels the left (source) and right (drain) contacts (i.e., $i=L,R$), and $\theta_{i}$ is the magnetization angle with respect to $z$-axis. We will designate the majority spin as `spin up' and the minority spins as `spin down'. $k_{i}^{\uparrow}=[2m_{x}(\epsilon-\epsilon_{y}-\epsilon_{j}-E_{i}^{\uparrow})]^{1/2}/\hbar$ is the wave-vector in the contact $i$, and the energy of the majority band edge is $E_{i}^{\uparrow}$. A similar identity holds for $k_{i}^{\downarrow}$ and $E_{i}^{\downarrow}$. The difference between the energies of the minority and the majority spin band edges is the exchange splitting in the ferromagnet $\Delta_{s} = E_{i}^{\downarrow} - E_{i}^{\uparrow}$. The Fermi energy (equivalently, the electrochemical potential) in contact $i$ 
is designated $\epsilon_{F}^{i}$. \textcolor{black}{The energy bandwidth of occupied states in HMF is defined as $E_{w}=\epsilon_{F}^{i}-E_{i}^{\uparrow}$ where $E_{i}^{\uparrow}$ is the energy of the majority spin band. }

In the ballistic case, states in the device are filled and emptied through the contacts. Conventionally, they can be defined as the filling and emptying functions (analogous to the in-scattering and out-scattering functions in the case of scattering \cite{datta97}),
\small
\begin{eqnarray}
\nonumber
\Sigma_{i}^{in}(\epsilon_{x})|_{\epsilon_{y},\epsilon_{j}}&=&f_{0}\left(\epsilon_{x}+\epsilon_{y}+\epsilon_{j}-\epsilon_{F}^{i}\right)\Gamma_{i}(\epsilon_{x})\\
\Sigma_{i}^{out}(\epsilon_{x})|_{\epsilon_{y},\epsilon_{j}}&=&\left[1-f_{0}\left(\epsilon_{x}+\epsilon_{y}+\epsilon_{j}-\epsilon_{F}^{i}\right)\right]\Gamma_{i}(\epsilon_{x})
\end{eqnarray}
\normalsize   
where $\Gamma_{i}=i[\Sigma_{i}-\Sigma_{i}^{\dagger}]$ is the broadening functions of the respective contact $i=L,R$. The electron and hole correlation functions are defined as
\small
\begin{eqnarray}
G^{n,p}(\epsilon_{x})|_{\epsilon_{y},\epsilon_{j}}=G(\epsilon_{x})\Sigma_{c}^{in,out}(\epsilon_{x})|_{\epsilon_{y},\epsilon_{j}}G(\epsilon_{x})^{\dagger}
\end{eqnarray}
\normalsize   
where $\Sigma_{c}^{in,out}=\Sigma_{L}^{in,out}+\Sigma_{R}^{in,out}$. The transverse modes can be summed over and we obtain the aggregated electron correlation function $\tilde{G}^{n}$,
\small
\begin{eqnarray}
\tilde{G}^{n}(\epsilon_{x})=G(\epsilon_{x})\tilde{\Sigma}_{c}^{in}(\epsilon_{x})G(\epsilon_{x})^{\dagger}
\end{eqnarray}
\normalsize   
with the aggregated filling function defined as
\small
\begin{eqnarray}
\tilde{\Sigma}_{c}^{in}(\epsilon_{x})=\sum_{j}F\left(\epsilon_{x}+\epsilon_{j}-\epsilon_{F}^{i}\right)\Gamma_{i}(\epsilon_{x})
\end{eqnarray}
\normalsize   
where $F(\epsilon_{x}+\epsilon_{j}-\epsilon_{F}^{i})$ is the Fermi Dirac integral of order $-\frac{1}{2}$ \cite{halen85}. The diagonal elements of $\tilde{G}^{n}(\epsilon_{x})$ are related to the charge spectral density at energy $\epsilon_{x}$. Once the total charge density is evaluated, the electrostatic potential $V_{1}$ can be obtained using Poisson equation self-consistently.

To include the effect of spin relaxation, we have to modify the Green's function in 
Eq.~(\ref{gfunction}). 
Spin relaxation processes arise from the interaction of spins of free carriers with the spins of localized electrons, e.g. in the `magnetically dead layer'. The Heisenberg Hamiltonian for the spin interaction is
\begin{eqnarray}
H_{I}=J \mathbf{s}\cdot \mathbf{S} .
\label{HeisenHam}
\end{eqnarray}
where the spin operators for free electrons are $\textbf{s}$ and those for localized electrons are $\textbf{S}$, all in units of $\hbar$. For spin=1/2 these operators are related to the Pauli matrices $\textbf{s}= \mathbf{\sigma} /2$. The interaction energy is given by $J$.

Assuming that the localized electrons are numerous, they thus form a reservoir causing an incoherent evolution of the free carrier spins. The state of the reservoir is described by its density matrix. For the case of spin=1/2 reservoir, it has the form
\begin{eqnarray}
\rho = \left( 
\begin{array}{cc}
F_u & \Delta \\
\Delta^* & F_d
\end{array}
\right)
\label{resdensmat}
\end{eqnarray}
where the spin-up and spin-down occupation numbers are $F_{u}$ and $F_{d}$ (such that $F_{u}+F_{d}=1$).

In this paper, we assume that spin relaxation processes are elastic (do not change the energy of free carriers). In the self-consistent Born approximation \cite{mahan90}, one can express the in- and out-scattering self-energy as a function of the electron and hole correlation functions \cite{datta04,yanik07},
\begin{eqnarray}
\nonumber
\Sigma^{in}_{s,ij}(\epsilon)&=& \gamma(\epsilon) \Phi^{n}_{ijkl} G^{n}_{kl}(\epsilon), \\
\Sigma^{out}_{s,ij}(\epsilon)&=& \gamma(\epsilon) \Phi^{p}_{ijkl} G^{p}_{kl}(\epsilon),
\label{sigmascat_orig}
\end{eqnarray}
where $\gamma(\epsilon)$ is the quantity with the dimension of 
energy-squared proportional to the relaxation rate, which depends on the number of localized spins and the interaction energy, $\Phi^{n/p}$ are the four-index tensors which provides a mapping between electron/hole correlation functions with the in/out-scattering functions. They are products of spin operators and the reservoir density matrix (\ref{resdensmat}). Their specific form and derivations are provided in Appendix \ref{sec:appen}. 

In the case of spin=1/2 reservoir and diagonal density matrix
($\Delta=0$) the explicit form can be derived 
\begin{widetext}
\small
\begin{eqnarray}
\nonumber
\Sigma_{s}^{in}(\epsilon_{x})|_{\epsilon_{y},\epsilon_{j}}&=&
{\gamma(\epsilon_{x})}
\left[
\begin{array}{cc}
F_{u}G^{n}_{\downarrow\downarrow}(\epsilon_{x})+\frac{1}{4}G^{n}_{\uparrow\uparrow}(\epsilon_{x}) & -\frac{1}{4}G^{n}_{\uparrow\downarrow}(\epsilon_{x})\\
-\frac{1}{4}G^{n}_{\downarrow\uparrow}(\epsilon_{x}) & F_{d}G^{n}_{\uparrow\uparrow}(\epsilon_{x})+\frac{1}{4}G^{n}_{\downarrow\downarrow}(\epsilon_{x})
\end{array}
\right]_{\epsilon_{y},\epsilon_{j}}\\
\Sigma_{s}^{out}(\epsilon_{x})|_{\epsilon_{y},\epsilon_{j}}&=&
{\gamma(\epsilon_{x})}
\left[
\begin{array}{cc}
F_{d}G^{p}_{\downarrow\downarrow}(\epsilon_{x})+\frac{1}{4}G^{p}_{\uparrow\uparrow}(\epsilon_{x}) & -\frac{1}{4}G^{p}_{\uparrow\downarrow}(\epsilon_{x})\\
-\frac{1}{4}G^{p}_{\downarrow\uparrow}(\epsilon_{x}) & F_{u}G^{p}_{\uparrow\uparrow}(\epsilon_{x})+\frac{1}{4}G^{p}_{\downarrow\downarrow}(\epsilon_{x})
\end{array}
\right]_{\epsilon_{y},\epsilon_{j}}
\label{sigmascat}
\end{eqnarray}
\normalsize 
\end{widetext}  

\textcolor{black}{The broadening function due to scattering is given by;
\small
\begin{eqnarray}
\Gamma_{s}(\epsilon_{x})|_{\epsilon_{y},\epsilon_{j}}=\left[\Sigma_{s}^{in}(\epsilon_{x})|_{\epsilon_{y},\epsilon_{j}}+\Sigma_{s}^{out}(\epsilon_{x})|_{\epsilon_{y},\epsilon_{j}}\right]
\end{eqnarray}
\normalsize   
If we assumed there is no coupling between the different $\epsilon_{y}$ and $\epsilon_{j}$ modes,
\small
\begin{eqnarray}
\Sigma_{s}(\epsilon_{x})|_{\epsilon_{y},\epsilon_{j}}=\frac{1}{2\pi}\int\frac{\Gamma_{s}(\epsilon_{x}')|_{\epsilon_{y},\epsilon_{j}}}{\epsilon_{x}'-\epsilon_{x}}d\epsilon_{x}'-i\frac{\Gamma_{s}(\epsilon_{x})|_{\epsilon_{y},\epsilon_{j}}}{2}
\label{hilberts}
\end{eqnarray}
\normalsize   
The non-coherent Green's function for a particular transverse mode ($\epsilon_{y}$,$\epsilon_{j}$) is given by,
\small
\begin{eqnarray}
G(\epsilon_{x})|_{\epsilon_{y},\epsilon_{j}}=\left[(\epsilon_{x}) I-H-V_{1}-\Sigma_{c}(\epsilon_{x})-\Sigma_{s}(\epsilon_{x})|_{\epsilon_{y},\epsilon_{j}}\right]^{-1}
\end{eqnarray}
\normalsize   
whereas the electron and hole correlation functions are defined as;
\small
\begin{eqnarray}
\nonumber
&&G^{n,p}(\epsilon_{x})|_{\epsilon_{y},\epsilon_{j}}=\\
&&G(\epsilon_{x})\left[\Sigma_{c}^{in,out}(\epsilon_{x})|_{\epsilon_{y},\epsilon_{j}}+\Sigma_{s}^{in,out}(\epsilon_{x})|_{\epsilon_{y},\epsilon_{j}}\right]G(\epsilon_{x})^{\dagger}
\end{eqnarray}
\normalsize   
The solutions are sought by solving the set of functions $\Sigma_{s}(\epsilon_{x})|_{\epsilon_{y},\epsilon_{j}}$, $G^{n,p}(\epsilon_{x})|_{\epsilon_{y},\epsilon_{j}}$ and $G(\epsilon_{x})|_{\epsilon_{y},\epsilon_{j}}$ self-consistently for each energy $\epsilon_{x}$ and modes ($\epsilon_{y},\epsilon_{j}$). This iterative process makes it numerically prohibitive to solving realistic transport problems.}

\textcolor{black}{In this work, we shall introduce some simplifications to make the numerics more tractable. Firstly, we assumed that the relaxation rate $\gamma$ is energy independent. Under the condition where the impurity spin state is uncorrelated and with equal up and down spin occupation probabilty i.e. $F_{u}$=$F_{d}$=$\frac{1}{2}$, it can also be shown that the function $\Phi^{n}$=$\Phi^{p}$=$\Phi$ \cite{yanik07}. Henceforth, the broadening due to scattering is given by;
\small
\begin{eqnarray}
\nonumber
\Gamma_{s}(\epsilon_{x})|_{\epsilon_{y},\epsilon_{j}}&=&
\gamma \Phi A(\epsilon_x)
\equiv \Gamma_{s}(\epsilon_{x})
\label{gammaeqs}
\end{eqnarray}
\normalsize   
where $A$ is the local density of state,
\small
\begin{eqnarray}
A(\epsilon_{x}) = i\left[
G(\epsilon_{x})-G(\epsilon_{x})^{\dagger}
\right]
= G^{n}(\epsilon_{x})+G^{p}(\epsilon_{x})
\end{eqnarray}
\normalsize   
We had assumed in Eq.~(\ref{gammaeqs}) that the local density of states for each transverse modes ($\epsilon_{y}$, $\epsilon_{j}$) is the same and is only dependent on the longitudinal energy $\epsilon_{x}$. Similarly, the scattering self energy $\Sigma_{s}(\epsilon_{x})$ (computed using Eq.~(\ref{hilberts})) and Green's function $G(\epsilon_{x})$ are modes independent. The aggregated electron correlation function can then be computed self-consistently from,
\small
\begin{eqnarray}
\tilde{G}^{n}(\epsilon_{x})=G(\epsilon_{x})\left[\tilde{\Sigma}_{s}^{in}(\epsilon_{x})+\tilde{\Sigma}_{c}^{in}(\epsilon_{x})\right]G(\epsilon_{x})^{\dagger}
\end{eqnarray}
\normalsize   
where the aggregated in-scattering self-energy for a spin $\frac{1}{2}$ impurity is written as,
\small
\begin{eqnarray}
\tilde{\Sigma}_{s}^{in}(\epsilon_{x})=
{\gamma}
\left[
\begin{array}{cc}
F_{u}G^{n}_{\downarrow\downarrow}(\epsilon_{x})+\frac{1}{4}G^{n}_{\uparrow\uparrow}(\epsilon_{x}) & -\frac{1}{4}G^{n}_{\uparrow\downarrow}(\epsilon_{x})\\
-\frac{1}{4}G^{n}_{\downarrow\uparrow}(\epsilon_{x}) & F_{d}G^{n}_{\uparrow\uparrow}(\epsilon_{x})+\frac{1}{4}G^{n}_{\downarrow\downarrow}(\epsilon_{x})
\end{array}
\right]
\end{eqnarray}
\normalsize 
The scattering strength of a layer of impurities is defined by the product $\gamma a$,
where $a$ is the lattice spacing, which physically speaking is the thickness of the interfacial layer in our study.
Current is calculated from the self-consistent 
solution of the above equations for any terminal $i$
\small
\begin{eqnarray}
I_{i}=\int_{-\infty}^{\infty}I(\epsilon_{x})d\epsilon_{x}
\end{eqnarray}
\normalsize   
where $I(\epsilon_{x})$ is defined as
\small
\begin{eqnarray}
I(\epsilon_{x})=\frac{q}{h} Tr
\left[\tilde{\Sigma}_{i}^{in}(\epsilon_{x})A(\epsilon_{x})-\Gamma_{i}(\epsilon_{x})\tilde{G}^{n}(\epsilon_{x})\right]
\end{eqnarray}
\normalsize   
The formalism described in this section allows us to combine the description of quantum transport in the device with the incoherent processes of spin relaxation.}

\section{\label{sec:coherent}Spin-MOSFET: Coherent Regime}

\begin{figure}[t]
\centering
\scalebox{0.32}[0.32]{\includegraphics*[viewport=2 16 782 570]{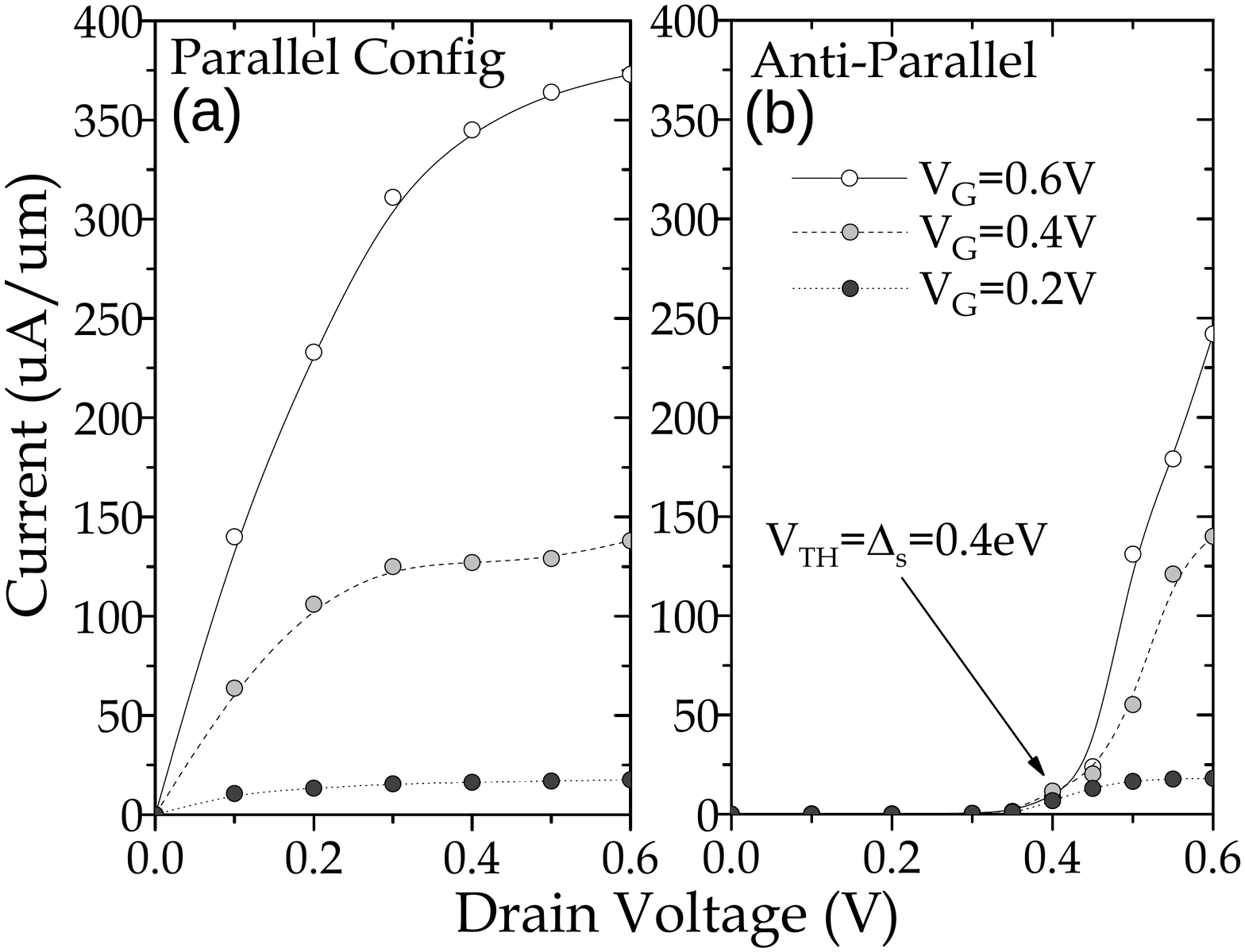}}
\caption{\footnotesize (a)  Current-Voltage characteristics for spin-MOSFET with the contacts magnetization in parallel configuration plotted for $V_{G}=0.2,0.4,0.6V$. (b)  Current-Voltage characteristics for spin-MOSFET with the contacts magnetization in anti-parallel configuration plotted for $V_{G}=0.2,0.4,0.6V$. }
\label{coherentMR3}
\end{figure}

\begin{figure}[t]
\centering
\scalebox{0.32}[0.32]{\includegraphics*[viewport=2 9 782 570]{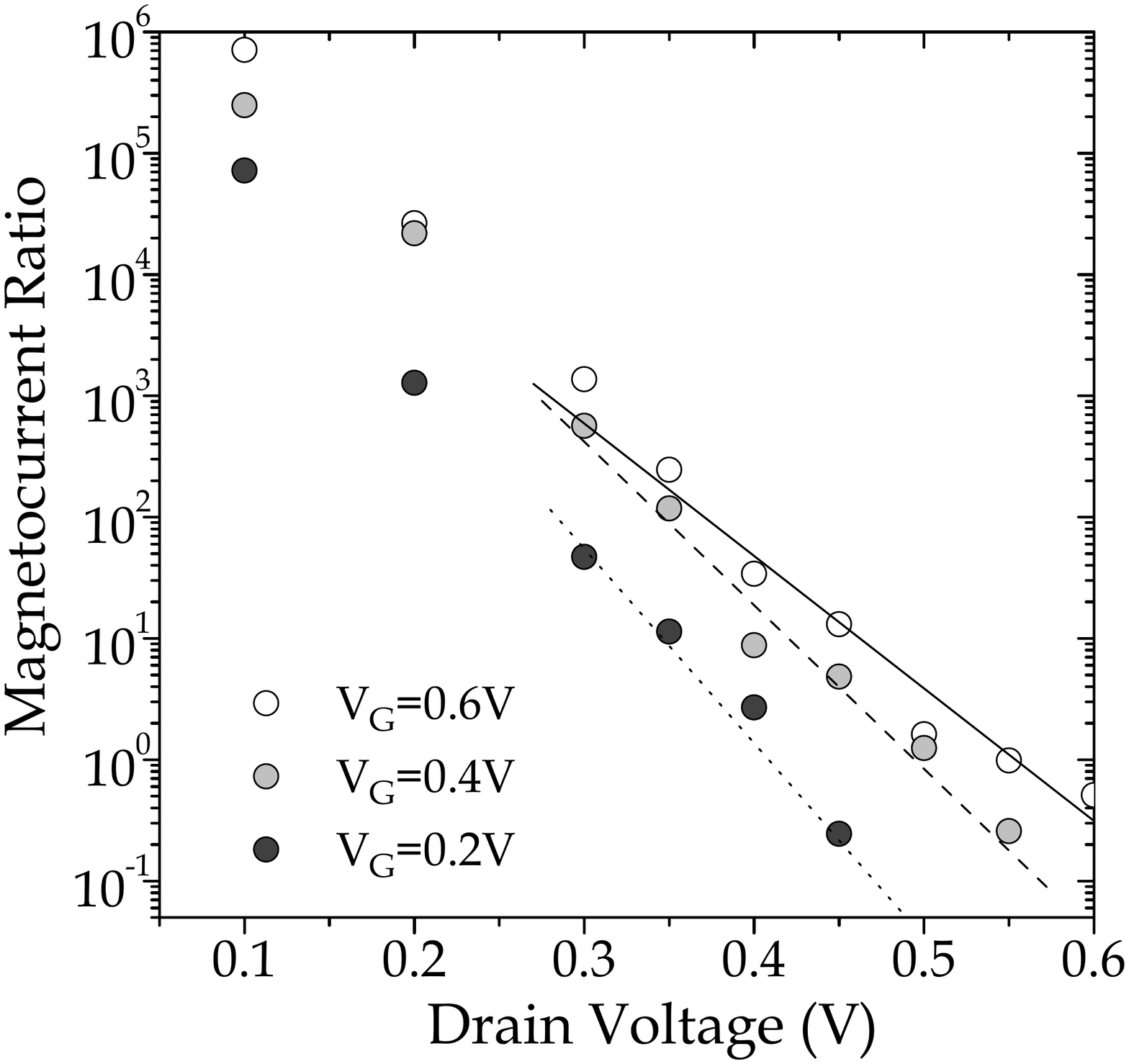}}
\caption{\footnotesize MR ratio, $MR=(I_{P}-I_{AP})/I_{AP}$, plotted for $V_{G}=0.2,0.4,0.6V$. The lines are the least square fitted to data points with $V_{D}$$\geq$$0.3V$ for each $V_{G}$. }
\label{MRcoherent}
\end{figure}

\begin{figure}[t]
\centering
\scalebox{0.32}[0.32]{\includegraphics*[viewport=2 41 780 580]{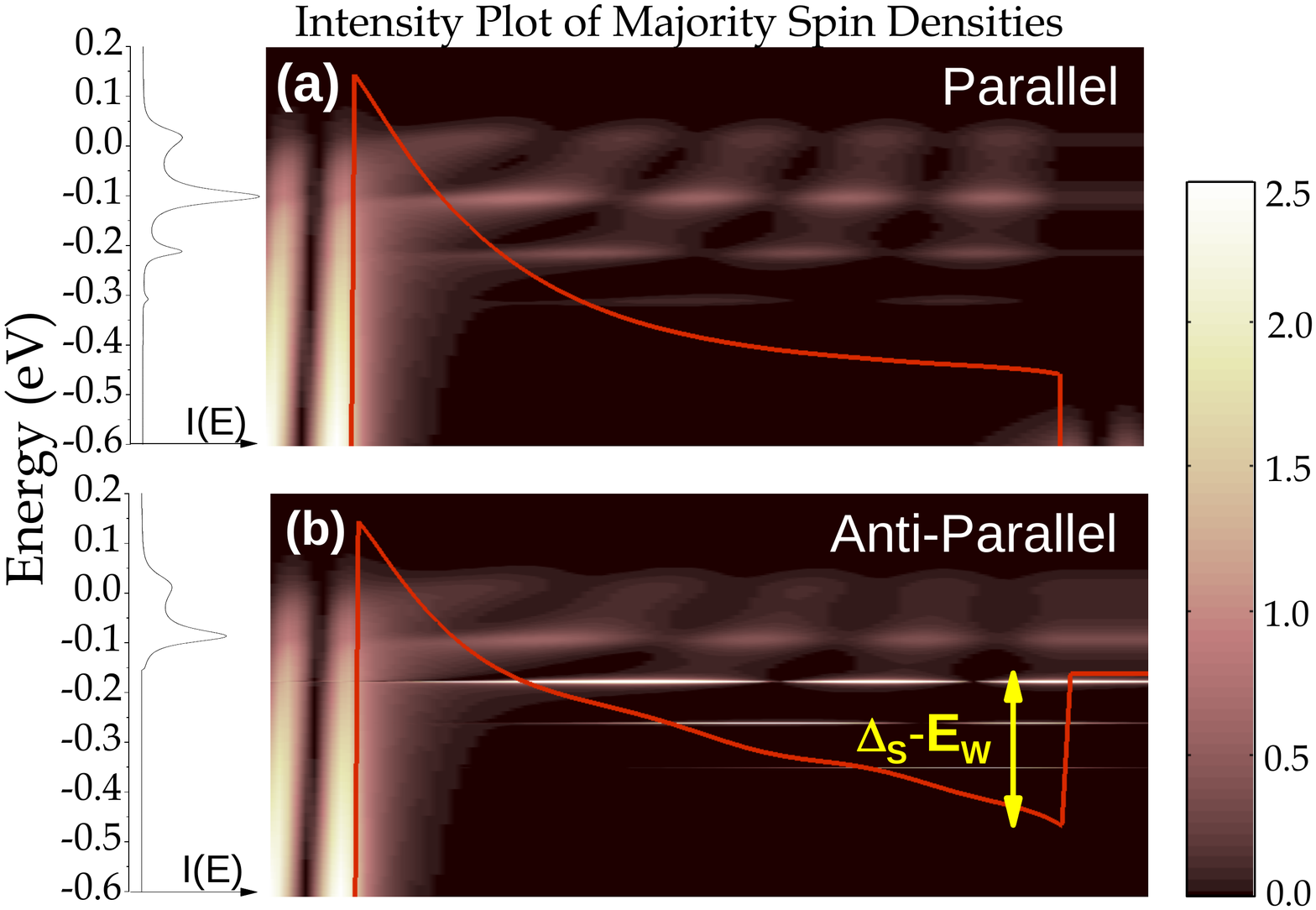}}
\caption{\footnotesize (a) Intensity plot of the majority spin density $D_{u}(\vec{r})$ (where it is scaled to present color contrast i.e. $D_{u}(\vec{r})^{0.8}$) for the case where magnetization is in (a) parallel and (b) anti-parallel configuration respectively. The terminal bias for both cases are $V_{G}=0.6V$ and $V_{D}=0.6V$. The energy-resolved current is plotted to the left. }
\label{coherentplot}
\end{figure}

We consider a spin MOSFET with a double-gated structure as depicted in Fig.\ref{deviceschematic} with a $3nm$ thin film Si channel. The channel length is $12nm$ with gate oxide thickness of $1nm$. Due to the strong body confinement, electrons predominantly occupy the doubly degenerate valleys along $\mathbf{k}=(0,0,1)$. Its energy dispersion can be described within the effective mass approximation with a longitudinal mass of $0.91m_{0}$ ($m_{z}$) and a transverse mass of $0.19m_{0}$ ($m_{x}$, $m_{y}$). HMFs are employed for the source and drain contacts. 

\textcolor{black}{Despite optimistic theoretical projections, engineering half metallic interfaces with semiconductors is still in its early stages of development \cite{galanakis02}. For the purpose of this work, we shall assume some reasonable material parameters for HMF to be used for our theoretical calculations. Heusler alloys of type X-MnSb have minority spin bands with energy gap ranging from $0.5eV$ to $1eV$ \cite{galanakis02}. Therefore, we assumed that the HMF's minority spin band energy minimum to be $~0.4eV$ above the metal Fermi energy. This value coincides with that of NiMnSb \cite{wijs01} based on a first principle calculation. The metallic nature of HFM is due to the electronic states occupying a large energy bandwidth, $E_{w}$, which is the energy of the conduction band minimum of majority spin band from Fermi energy. In this work, we arbitrarily set the metal's occupied bandwidth to $E_{w}$=$2eV$. We also assumed that the majority and minority spin bands in HFM has the same transport mass as Si. The choice of different transport mass in HMF would essentially introduce more reflections at the interfaces, which can also be compensated by a larger $E_{w}$. However, in the general case (e.g. Fe), the minority and majority spin bands have to be modeled differently \cite{bagrets02}.}

The current-voltage characteristics of spin-MOSFETs for the case of parallel and anti-parallel magnetization configurations in the coherent regime are plotted in Fig.\ref{coherentMR3}(a) and (b), respectively. \textcolor{black}{They exhibit similar current-voltage characteristics, except that the anti-parallel magnetization configurations exhibits a drain offset voltage by an amount of $\Delta_{s}$-$E_{w}$.} This is due to the potential blockade of the majority spin at the drain HMF contact in the anti-parallel configuration (see Fig.\ref{coherentplot}(b)). Fig.\ref{coherentplot} is an intensity plot of the majority carrier density of spin MOSFETs for the case of parallel and anti-parallel magnetization configurations. Also shown in Fig.\ref{coherentplot} is the energy-resolved current (on the left). Oscillations in the energy resolved current for the parallel case are signatures of tunneling through a barrier, e.g. commonly observed in a Fowler Nordheim tunneling through $Si|SiO_{2}|Si$ sandwiched structure \cite{gundlach66}. In the anti-parallel case (Fig.\ref{coherentplot}(b)), the potential barrier at the drain HMF contact permits transmission of the majority spin only when carrier energy is greater than $\Delta_{s}$. Resonance states due to lateral confinement are observed when the carrier energy is less than $\Delta_{s}$, and these states will not contribute any current. These resonance states results in the oscillatory behavior in the derivatives of the potential profile in the anti-parallel configuration. \textcolor{black}{It is also numerically challenging to resolve these states in the energy domain due to the relatively fine linewidth in these strongly localised resonance levels i.e. a numerical challenge faced similarly in modeling of resonant tunneling diode. A fine gridding is employed to resolve these resonance states so that the charge density can be more reliably computed.}

The magnetocurrent ratio ($MR$) is defined to be $MR$$=$$(I_{P}$$-$$I_{AP})$$/$$I_{AP}$. It serves to quantify the magneto-resistance difference between the parallel and anti-parallel magnetization configuration states of the device. Since the device in the anti-parallel state has an apparent drain offset voltage of $\Delta_{s}$-$E_{w}$, $I_{AP}$ will generally be less than $I_{P}$. Fig.\ref{MRcoherent} plots the MR of the spin-MOSFET under different biasing conditions. The MR follow an approximately \textcolor{black}{exponential relationship} with $V_{D}$ at a given $V_{G}$ bias. We can also see that the MR begins to diminished ($MR$$<$$10\%$) when the $V_{D}$ in the anti-parallel configuration approaches the current saturation condition i.e. $V_{D}=V_{sat}$. Consequently, there is a rightward shifts of the MR versus $V_{D}$ curve when the gate bias $V_{G}$ increases, since $V_{sat}$ increases with increasing $V_{G}$.

\section{\label{sec:incoherent}Spin-MOSFET: Incoherent Regime}

\begin{figure}[t]
\centering
\scalebox{0.45}[0.45]{\includegraphics*[viewport=45 125 570 600]{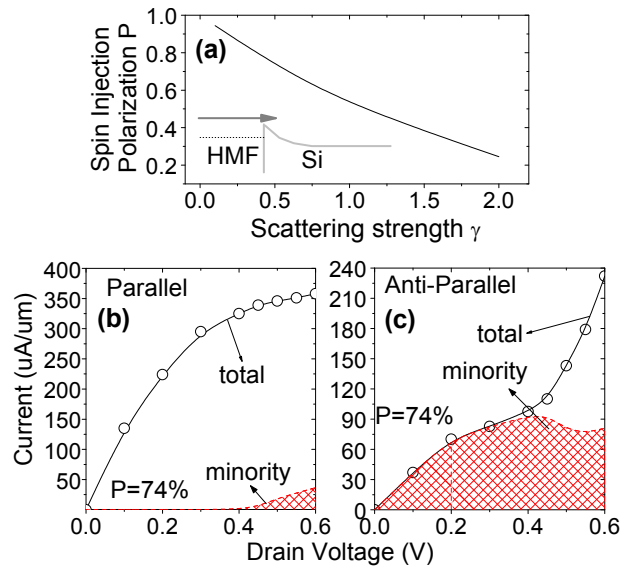}}
\caption{\footnotesize (a) Spin polarization $P$ in the channel as a function of the interface scattering strength characterized by $\gamma a$, for spin injection through a Schottky barrier in a HMF/Si structure. (b) Current-voltage characteristics for spin MOSFET with the contacts magnetization in parallel configuration plotted for $V_{G}$=$0.6V$. All spin currents are measured at the drain. (c) Current-voltage characteristics for spin-MOSFET with the contacts magnetization in anti-parallel configuration plotted for $V_{G}$=$0.6V$. For both cases, the spin relaxation strength at the HMF/semiconductor interfaces on both the detector and injector sides are characterized by $\gamma a=0.5eV^{2}nm$. The minority spin current is plotted as dashed curve. }
\label{incoherent2}
\end{figure}  

Spin exchange scattering processes between the tunneling electron and the localized spin impurities at the HMF/Si interfaces are responsible for the incoherent nature of the electron transport. These localized spin impurities lead to decoherence of the electronic spins states. In our model, we assume that there are external reservoirs constantly maintaining these localized spin impurities in a state of equilibrium with a random polarization of spin. For a spin=1/2 case, the completely unpolarized states corresponds to the density matrix with $F_{u}$=$F_{d}$=$0.5$ and $\Delta$=$0$ \cite{datta04}, \footnote{For spin impurities such as nuclear spins, these spin flip interactions can easily overwhelm the external forces and polarize the impurities so that they are no longer effective in flipping electronic spins}. 

\begin{figure*}[t]
\centering
\scalebox{0.6}[0.6]{\includegraphics*[viewport=17 310 720 580]{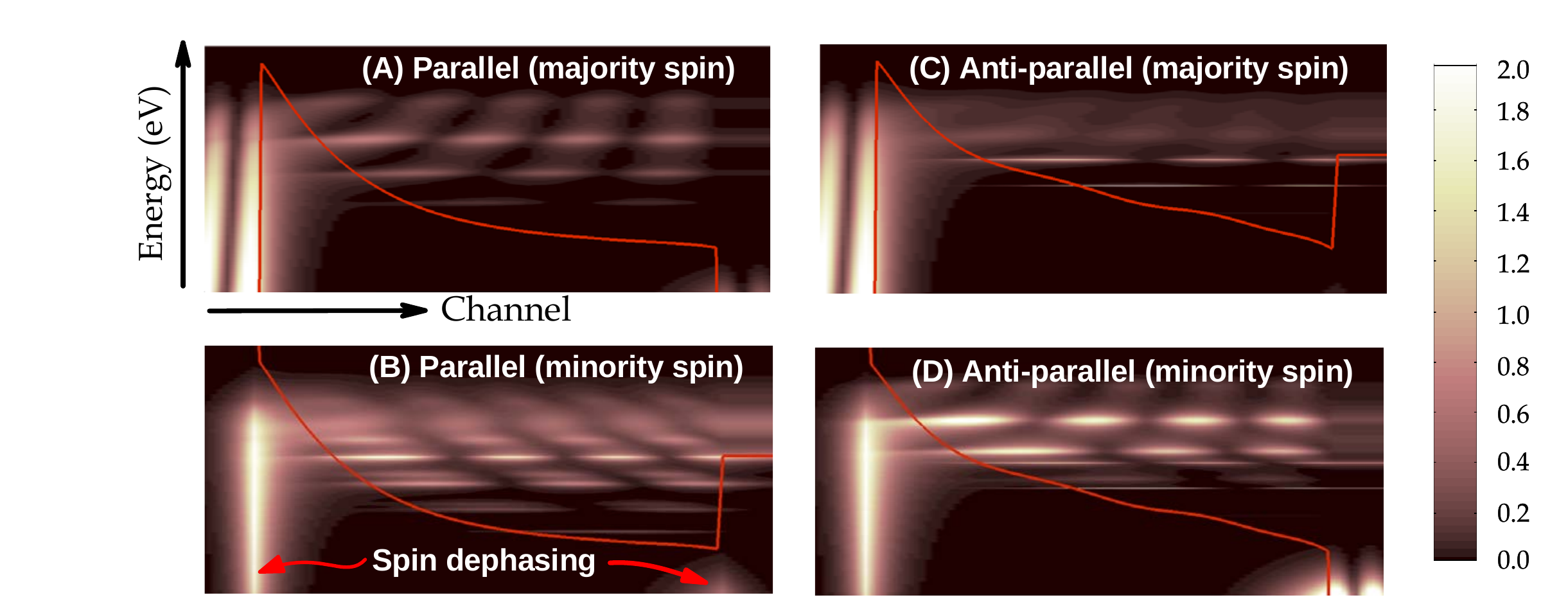}}
\caption{\footnotesize Intensity plot of the (a) majority spin density $n_{u}(\vec{r})$ and (b) minority spin density $n_{d}(\vec{r})$ (where it is scaled to present color contrast i.e. $n_{u}(\vec{r})^{0.8}$ and $10\times n_{d}(\vec{r})^{0.8}$ respectively) for the case where magnetization is in parallel. Similar plots for the majority and minority spin density for the anti-parallel configuration in (c) and (d) respectively. The terminal bias for both cases are $V_{G}$=$V_{D}$=$0.6V$.}
\label{incoherentplot2}
\end{figure*}  

\begin{figure}[t]
\centering
\scalebox{0.35}[0.35]{\includegraphics*[viewport=50 20 900 600]{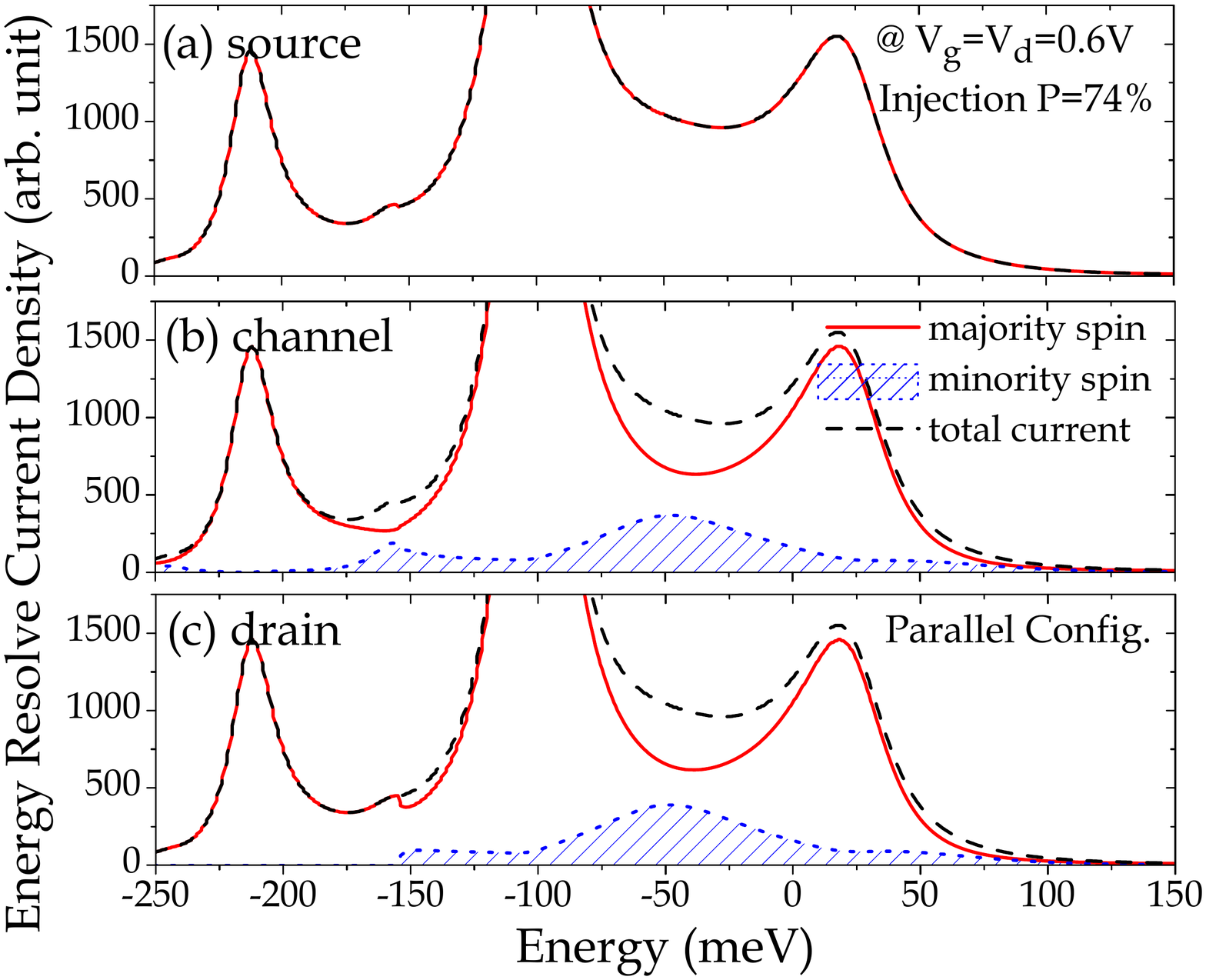}}
\caption{\footnotesize Spin current spectral density vs. carrier energy, for the parallel directions of magnetization of source and drain. Total current - black dashed line, majority spin - red solid line, minority spin - blue shaded area. Top plot - at the source, middle- plot - in the center of the channel, bottom plot - at the drain.  }
\label{currentscat_p}
\end{figure}  
\begin{figure}[t]
\centering
\scalebox{0.35}[0.35]{\includegraphics*[viewport=50 20 900 600]{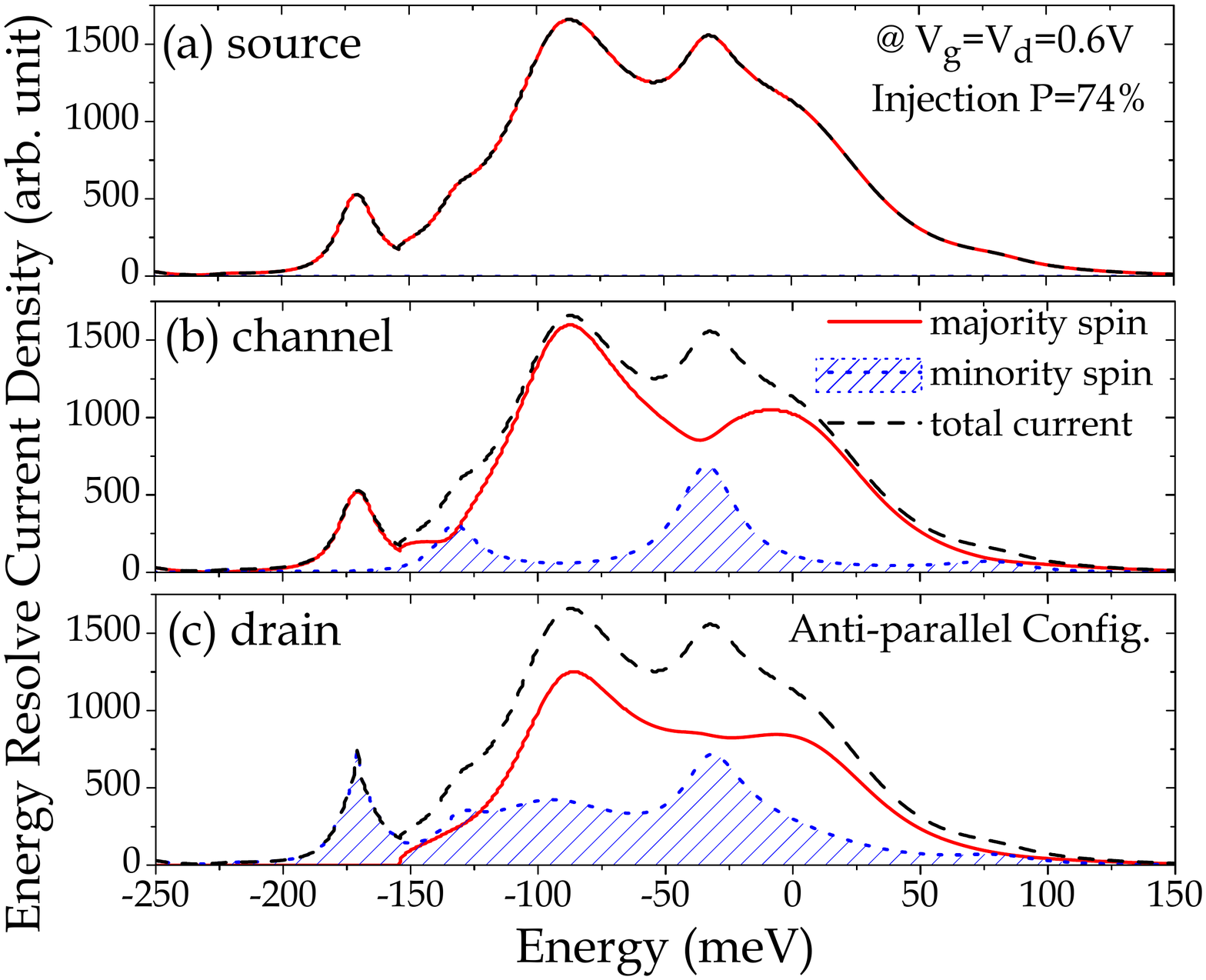}}
\caption{\footnotesize Same as in Fig.~\ref{currentscat_p}, for the anti-parallel directions of magnetization of source and drain. }
\label{currentscat_ap}
\end{figure} 
As discussed in Sec II, coupling between the number of available electrons/holes ($[G^{n}]/[G^{p}]$) for a particular state, and the in/out-flow ($[\Sigma^{in}_{s}]$/$[\Sigma^{out}_{s}]$) to/from that state is related through a four-index scattering tensors $\Phi$. The scattering strength, $\gamma a$, is assumed to be energy independent. It is proportional to $n_{I}$, the impurity concentration and to $\left\langle J^{2}\right\rangle$, the averaged interaction energy of the impurity layer (see Appendix \ref{sec:appen}). These parameters can be mapped to 
available experimental data, and they are the only parameters used to characterize the spin relaxation of the impurity layer. Fig.\ref{incoherent2}(a) shows the spin polarization $P$ in the channel due to spin injection from a HMF contact into a long-channel semiconductor. $P$ is defined as;
\small
\begin{eqnarray}
P=\frac{I_{\uparrow}-I_{\downarrow}}{I_{\uparrow}+I_{\downarrow}}
\end{eqnarray}
\normalsize 
Our model is capable of capturing the non-ideal spin injection of a HMF due to the presence of a magnetically dead layer, by tuning the value of $\gamma a$ (see Fig.\ref{incoherent2}a). The spin polarization in the channel decreases with the increase of spin relaxation strength in an approximately linear fashion.\\

Fig.\ref{incoherent2}b and \ref{incoherent2}c plots the drain current vs. drain voltages at $V_{G}$$=$$0.6V$ for both the parallel and anti-parallel configuration, in the presence of spin relaxation at the HMF/semiconductor interfaces on both the detector and injector sides, characterized by spin relaxation strength of $\gamma a = 0.5eV^{2}nm$. In contrast to the current-voltage characteristics in the coherent case (see Fig.\ref{coherentMR3}), the minority spin can now contribute to spin current, which significantly modifies the current-voltage characteristics in the anti-parallel configuration. This `leakage current' is facilitated through spin relaxation at the HMF/semiconductor interfaces. Spin relaxation renders the blocking (due to the potential barrier at the drain HMF) of the majority spin transmission ineffective. Majority carriers injected from the source can now undergo spin relaxation at the HMF/semiconductor interface at the drain side and become a minority spin. Therefore, substantial amount of minority spin current is registered in the anti-parallel configuration even when the drain bias is less than $\Delta_{s}$-$E_{w}$. 

Fig.\ref{incoherentplot2}(a)-(d) plots the majority and minority spin densities in the parallel and anti-parallel configuration at $V_{G}$$=$$V_{D}$$=$$0.6V$. Their respective energy resolved current is plotted in Fig.\ref{currentscat_p}-\ref{currentscat_ap}. In this set of calculations, the minority spin density is approximately an order of magnitude smaller than the majority spin density. In the intensity plot for the minority spin density, we observe signatures of spin flip process at the spin injector interface as a layer of high intensity minority spins. Majority spin was scattered into the minority spin states in the injector as evanescent states with finite probability to tunnel through the Schottky barrier into the channel. Therefore, HMF will not have perfect spin injection efficiency when spin relaxation mechanisms are incorporated into the model.

\textcolor{black}{Fig. \ref{currentscat_p} and \ref{currentscat_ap} plots the corresponding energy-resolved spin current density in each of the respective device regions i.e. (a) source, (b) channel and (c) drain, when the device is in the parallel and anti-parallel magnetization state respectively. Although the spin current is not conserved across the device, the charge current at each energy is conserved i.e. a property ensured by virtue of the self-consistent Born framework. From the energy resolved spin current characteristics, we can make the observation that more minority spin current is produced in the anti-parallel configuration than its parallel counterpart. }The energy resolved minority spin current in the anti-parallel configuration is characterized by multiple resonance peaks. This is in part due to the laterally confined majority spins in the channel where the minority spin current was derived from it through spin relaxation processes. 

\section{\label{sec:magnetoratio} Influence of Spin relaxation on MR Ratio}

\begin{figure}[t]
\centering
\scalebox{0.5}[0.5]{\includegraphics*[viewport=40 80 530 360]{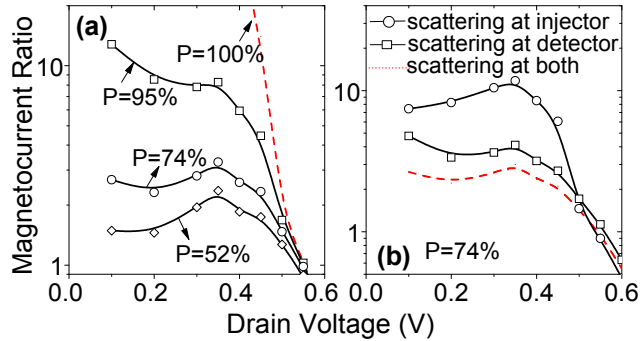}}
\caption{\footnotesize (a)  MR ratio $(I_{P}-I_{AP})/I_{AP}$ versus drain  biases for $V_{G}=0.6V$. The spin relaxation strength at the HMF/semiconductor interfaces on both the detector and injector sides are characterized by $\gamma a$=$0,0.1,0.5,1.0eV^{2}nm$. (b)  MR ratio versus drain  biases for $V_{G}=0.6V$ for spin relaxation at both detector and injector HMF/semiconductor interfaces (dashed line), at detector HMF/semiconductor interface only (solid line with square symbol) and at injector HMF/semiconductor interface only (solid line with circle symbol).  }
\label{JMRincoherent2}
\end{figure}

In this section, we discuss MR ratio in the presence of spin relaxation. Fig.\ref{JMRincoherent2}(a) shows the MR ratio versus the drain voltage bias at $V_{G}$$=$$0.6V$ in the presence of spin relaxation at the HMF/semiconductor interfaces on both the detector and injector sides, characterized by spin relaxation strength of $\gamma a = 0-1eV^{2}nm$. As evident in Fig.\ref{JMRincoherent2}(a), the increasing spin relaxation strength at the HMF/semiconductor interfaces results in decreasing MR ratio over the whole drain voltages sweep, especially for drain biases less than $\Delta_{s}$-$E_{w}$. Spin relaxation process enhance the probability of conduction through the minority spin channel which consequently smears the distinction between transport in the parallel and anti-parallel configuration, resulting in diminishing MR ratio. \textcolor{black}{Our analysis in this work focuses only on HMF contacts. The impact of interfacial spin dephasing on normal ferromagnetic contacts could have intrinsically different dependence on the spin dephasing processes. In the ohmic regime, one could model the interfacial spin dephasing process with a resistor element ($r_{\gamma}$) between the majority and minority spin channels at each contact interfaces. A simple electrical analysis of such a setup for spin-MOSFETs with either HFM or normal ferromagnetic contacts shows that they have different dependence on $r_{\gamma}$ (see Appendix \ref{sec:appenb}).}


\begin{figure}[t]
\centering
\scalebox{0.5}[0.5]{\includegraphics*[viewport=50 280 550 580]{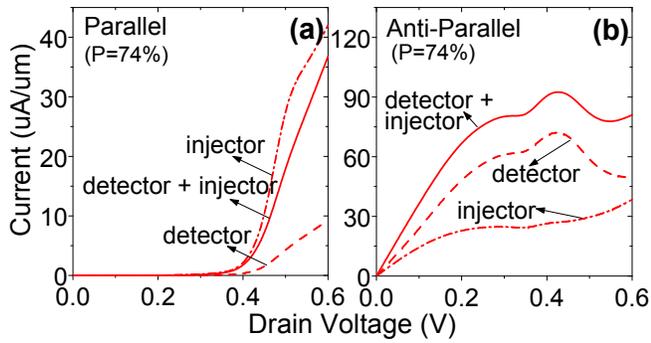}}
\caption{\footnotesize Minority spin current versus drain bias for $V_{G}$=$0.6V$ at (a) parallel and (b) anti-parallel configuration. The spin relaxation strength at the HMF/semiconductor interfaces on either/both the detector and injector sides are characterized by 
$\gamma a = 0.5eV^{2}nm$.
 We consider the situation where spin relaxation are at both detector and injector HMF/semiconductor interfaces (solid line), at detector HMF/semiconductor interface only (dotted line) and at injector HMF/semiconductor interface only (dash-dotted line). }
\label{interfacestudy}
\end{figure}

\textcolor{black}{Lastly, we studied the impact of spin relaxation at each of HMF/semiconductor interfaces. Fig.\ref{JMRincoherent2}(b) plots the MR ratio for spin relaxation at the HMF/semiconductor interfaces at either the injector or detector side. It is evident from Fig.\ref{JMRincoherent2}(b) that spin relaxation at the HMF/semiconductor interfaces at the detector side is more detrimental to the MR ratio than that due to the spin relaxation at the injector side, for drain bias less than $\Delta_{s}$-$E_{w}$. For each of these cases, Fig.\ref{interfacestudy} shows the minority spin current in the parallel and anti-parallel configurations at different drain biases. As discussed previously, the fractional contribution of minority spin current in the parallel case is relatively small and arises only when the drain bias is larger than $\Delta_{s}$-$E_{w}$. More minority spin current is produced in the anti-parallel case in this regime. From Fig.\ref{interfacestudy}(b), it is evident that more minority spin current is produced when spin relaxation at the HMF/semiconductor interfaces is at the detector side compared to when it is at the injector side. The consequence is that spin relaxation at the HMF/semiconductor interfaces at the detector side is more detrimental to the MR ratio, at least for drain bias less than $\Delta_{s}$-$E_{w}$. A plausible explanation to why spin relaxation at the detector side produces more minority spin current in anti-parallel case could be understood from Fig.\ref{incoherentplot2}(d). When $V_{D}<$$\Delta_{s}$-$E_{w}$, majority carrier undergoing spin relaxation at the detector interface will either be admitted into the drain as a minority spin or be reflected back into the channel as majority spin. Waves that backscatters at the detector interface will then be reflected back and undergoes another scattering event. This multiple scattering processes enhances the spin relaxation rate. This study reveals that proper HMF/semiconductor interface treatment at the detector side is more pertinent to achieving high MR ratio in spin MOSFET for operation regime where biasing conditions is less than $\Delta_{s}$-$E_{w}$. }

\textcolor{black}{We note that a spin MOSFET uses magnetoresistance caused by the change in spin transport through the device.
However in experimental relaizations, a large role is played by fringe magnetic fields near the ferromagnets which
might induce local Hall effect in the semiconductor channel at vicinity of the interface. This results in a spurious contribution to the magnetoresistance, as it was explained in \cite{hammar99,monzon00,wees00}. Measurements based on the precession of spins in the channel and the corresponding Hanle effect provide a rigorous method to detect the spin transport \cite{huang07,appelbaum07,lou07}. Practically, one should be able to minimize this spurious Hall effect via a proper geometrical setup of the spin injection device \cite{hammar00}. }


\section{\label{sec:conclude}Summary}

In this paper, we have simulated the operation of a spin MOSFET using the non-equilibrium Green's function approach. A large spin-splitting energy $\Delta_{s}$ for the HMF contacts is beneficial for achieving a large MR ratio. This is because the spin-splitting energy presents a potential barrier at the drain side in the anti-parallel configuration and blocks the transmission of majority spins, resulting in a larger drain `threshold voltage' of $~\Delta_{s}$-$E_{w}$ for the device in the anti-parallel configuration. We highlighted and explained the bias dependence of MR ratio of spin MOSFET. Next, we demonstrated the incorporation of spin relaxation at the HMF/semiconductor interfaces via scattering self-energies. Once spin relaxation is included in the model, minority spin current arises and it dominates the `leakage current' in the anti-parallel configuration even when the drain bias is less than $\Delta_{s}$-$E_{w}$. This substantially reduces the spin MOSFET MR ratio. Lastly, we studied the impact of the `magnetically dead layer' at the injector and at the detector sides in isolation. It was found that spin relaxation caused by the `magnetically dead layer' at detector side is more detrimental to the MR ratio. This can be attributed to the fact that spin relaxation due to the `magnetically dead layer' at the detector side is more efficient in flipping the majority spin to minority spin current.

\begin{acknowledgments}
T.L. and M.S.L. gratefully acknowledge support of the Nanoelectronic Research Initiative and the Network for Computational Nanotechnology for computational resources. They also thank Sayeef Salahuddin for careful proofreading and suggestions. D.E.N. is grateful to George Bourianoff and Paolo Gargini for stimulating discussions.
\end{acknowledgments}

\appendix
\section{\label{sec:appen} Derivation of Scattering Spin Tensors}

In this appendix we provide a simplified derivation and the 
explicit form in a particular case of spin=1/2 of the scattering tensors in 
Eqs.~(\ref{sigmascat}).
It follows earlier papers \cite{datta04,yanik07}, 
but is presented here for completeness sake.
In general, the scattering tensor is determined by the Hamiltonian 
of interaction with the reservoir $H_I$ as follows \cite{datta04}
\small
\begin{eqnarray}
\nonumber
\gamma \Phi^{n}_{ijkl} &=& \sum_{N_I} Tr \left[ \rho H_{Ilj} H_{Iik} \right], \\
\gamma \Phi^{p}_{ijkl} &=& \sum_{N_I} Tr \left[ \rho H_{Iik} H_{Ilj} \right],
\label{tensorsDatta}
\end{eqnarray}
\normalsize
where $\rho$ the density matrix of the reservoir spin, like in Eq.~(\ref{resdensmat}),
and $N_I$ is the number of reservoir modes. 
In our case, the Hamiltonian (\ref{HeisenHam})
corresponds to the spin-spin interaction.
The spin-dependent four-index tensors are thus
\small
\begin{eqnarray}
\Phi^{n}_{ijkl} &=& Tr \left[ \rho S^\alpha S^\beta  \right] s^\alpha_{lj} s^\beta_{ik} ,
\end{eqnarray}\normalsize
where indices $\alpha$ and $\beta$ run over the projections of the operators on Cartesian axes,
and summation over repeating index is implied.

And the prefactor contains the interaction energy $J$ 
averaged over the spectral range of reservoir modes resonant with the
transition (this averaging is designated by angled brackets), 
which according to \cite{yanik07} is,
\small
\begin{eqnarray}
\gamma = \sum_{N_I} J^2 \propto n_I \left\langle J^{2}\right\rangle,
\end{eqnarray}\normalsize
where $n_I$ is the area density of impurities in the layer and averaging is done
with the account of geometry of the 2D layer.

For the particular case of the spins of localized electrons (reservoir) equal
to 1/2, more explicit expressions can be obtained.
It is convenient to use the raising and lowering Pauli matrices
\small
\begin{eqnarray}
\sigma^{+}&=&\frac{1}{2}\left[\sigma^{x}+i\sigma^{y}\right], \\
\sigma^{-}&=&\frac{1}{2}\left[\sigma^{x}-i\sigma^{y}\right],
\end{eqnarray}\normalsize
The interaction Hamiltonian can be re-written as
\small
\begin{eqnarray}
H_{I}&=& J
\left[
\frac{\sigma^{+} S^{-}}{4} +
\frac{\sigma^{-} S^{+}}{4} +
\frac{\sigma^{z} S^{z}}{2}
\right].
\end{eqnarray}\normalsize
It is easy to evaluate the averages of the spin operators in the state of the reservoir
(\ref{resdensmat}). The scattering tensor becomes
\small
\begin{eqnarray}
\nonumber
\Phi^{n}_{ijkl} = 
(F_u + F_d) s^z_{lj} s^z_{ik} 
+ \Delta s^+_{lj} s^z_{ik} 
- \Delta s^z_{lj} s^+_{ik} \\
+ \Delta^* s^-_{lj} s^z_{ik} 
- \Delta^* s^z_{lj} s^-_{ik} 
+ F_d s^+_{lj} s^-_{ik} 
+ F_u s^-_{lj} s^+_{ik} ,
\end{eqnarray}\normalsize
and similarly for $\Phi^{p}$.

\section{\label{sec:appenb} Electrical Analysis of HMF/FM Contacts in Ohmic Regime}
\begin{figure}[htps]
\centering
\scalebox{0.6}[0.6]{\includegraphics*[viewport=50 556 520 790]{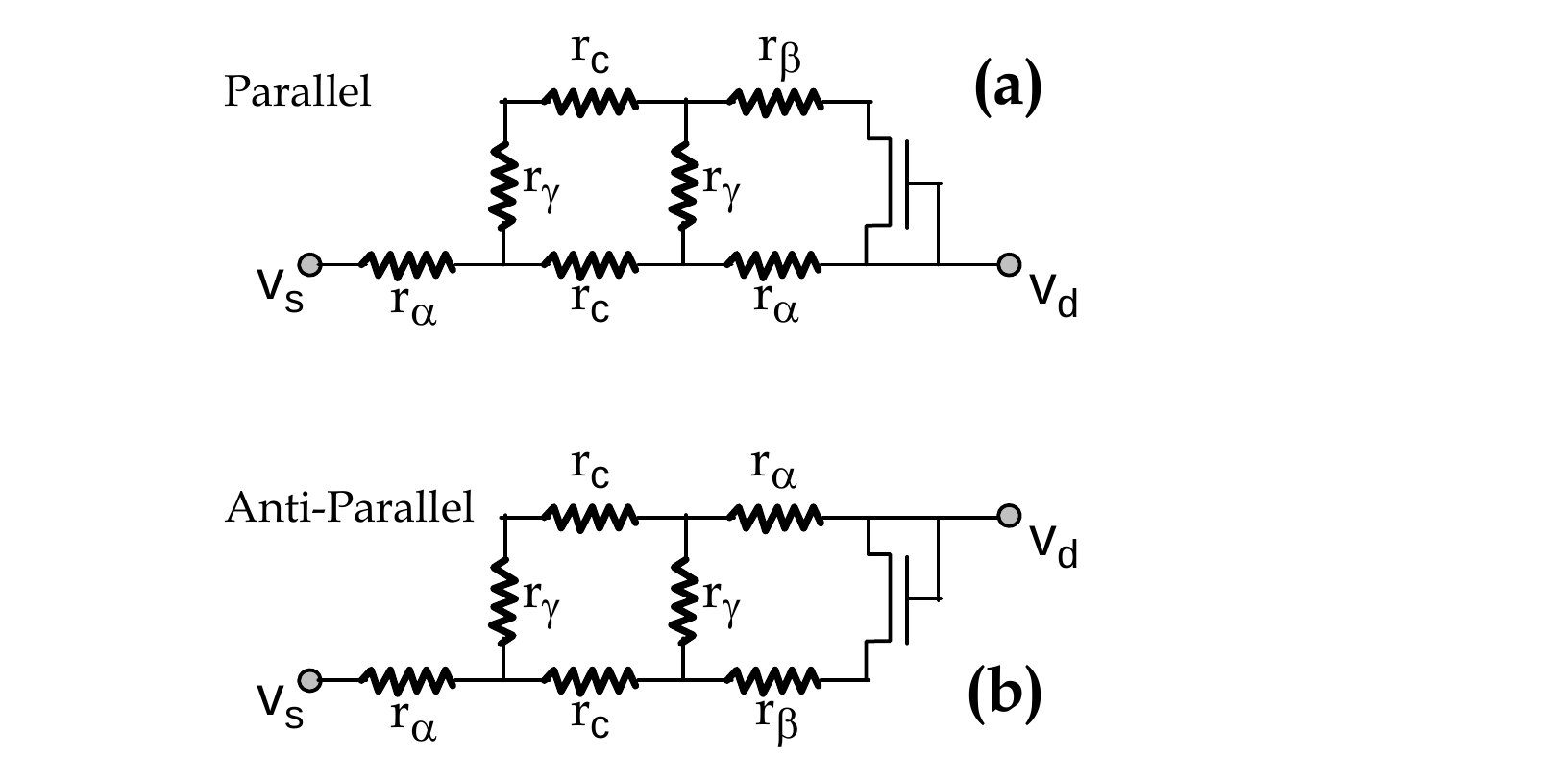}}
\caption{\footnotesize\slshape\sffamily (a) and (b) plot the equivalent circuit for the parallel and anti-parallel configuration of a spin MOSFET in the Ohmic regime where $r_{\alpha/\beta}$ is the resistance of the majority/minority spins in the HFM contacts, $r_{c}$ the channel resistance and $r_{\gamma}$ is the resistance related to spin dephasing. The transistor switch in the equivalent circuit model has a threshold voltage of $\Delta_{s}$-$E_{w}$. }
\label{schematic2}
\end{figure}  

\textcolor{black}{The magnetocurrent ratio is an important device metric which provides a measure of the degree of distinguishability between the parallel and anti-parallel states of the spin transistor through their measured current. The Julliere's description of the magnetocurrent ratio in terms of the available effective tunneling density of states proves to be inadequate \cite{maclaren97}, especially when there is interfacial spin flip scattering processes. A simple analysis using the electrical circuit equivalent in Fig.\ref{schematic2}(a) and (b) can offer more reliable insights. }\\

\textcolor{black}{Fig.\ref{schematic2}(a) and (b) plot the equivalent circuit for the parallel and anti-parallel configuration in the Ohmic regime where $r_{\alpha/\beta}$ ($g_{\alpha/\beta}$) is the resistance (conductance) of the majority/minority spins in the ferromagnetic contacts, $r_{c}$ ($g_{c}$) the channel resistance (conductance) and $r_{\gamma}$ ($g_{\gamma}$) is the resistance (conductance) related to spin dephasing. The transistor in the circuit is to model the effect of the minority/majority spin blocking due to the spin exchange energy at the drain HMF in the parallel/anti-parallel configuration. A simple Ohmic model can adequately explains the late turn on of the minority current in parallel configuration at $V_{D}$=$\Delta_{s}$-$E_{w}$ as depicted in Fig.\ref{incoherent2}. Recall that in the parallel state, no minority spin current can be detected at the HMF drain when $V_{D}$$<$$\Delta_{s}$-$E_{w}$, as the source injected states (with energy lesser than the source Fermi energy) admitted into the drain will either decay evanescently or revealed itself as resonant levels in the channel (see Fig.\ref{incoherentplot2}(d)). In addition, it also explains why the minority spin current in the anti-parallel configuration can leak effortlessly into the drain once the spin dephasing process opens up the minority spin channel through $g_{\gamma}$. }\\

\textcolor{black}{Consider the regime where the drain bias is less than $\Delta_{s}$-$E_{w}$, the respective resistance for the parallel and anti-parallel configuration are written as;
\small
\begin{eqnarray}
\nonumber
r_{P}&=&2r_{\alpha}+\frac{1}{2}\frac{r_{c}\left(r_{c}+2r_{\gamma}\right)}{r_{c}+r_{\gamma}}\\
r_{AP}&=&2r_{\alpha}+\frac{1}{2}\left(r_{c}+r_{\gamma}\right)
\end{eqnarray}
\normalsize 
which both converges to $2r_{\alpha}+r_{c}/2$ when the scattering is large i.e. $r_{\gamma}$$\ll$$r_{c}$, yielding a zero magnetocurrent ratio. Maximum magnetocurrent ratio is achieved when the scattering is minimal i.e. $r_{\gamma}$$\gg$$r_{c}$, with a large magnetocurrent ratio of the order $r_{\gamma}/r_{c}$. We can also derive similar expressions for normal ferromagnetic contacts;
\small
\begin{eqnarray}
\nonumber
r_{P}&=&\frac{2r_{\alpha}r_{\beta}}{r_{\tau}}+\frac{(2r_{\gamma}r_{\alpha}+r_{\tau} c)(2r_{\gamma}r_{\beta}+r_{\tau} r_{c})}{r_{\tau}(2r_{\gamma}r_{\beta}+r_{\tau}r_{c})+r_{\tau}(2r_{\gamma}r_{\alpha}+r_{\tau} r_{c})}\\
r_{AP}&=&\frac{2r_{\alpha}r_{\beta}}{r_{\tau}}+\frac{r_{\gamma}(r_{\alpha}+r_{\beta})+r_{\tau} r_{c}}{2r_{\tau}}
\end{eqnarray}
\normalsize 
where $r_{\tau}$$=$$r_{\alpha}$$+$$r_{\beta}$$+$$r_{\gamma}$. By making the assumption that the $r_{c}$$\ll$$r_{\gamma}$, we arrive at the result that the magnetocurrent ratio for the HMF and FM case varies with the spin dephasing conductance $g_{\gamma}$ according to ~$g_{\gamma}^{-1}$ and ~$g_{\gamma}^{-2}$ under these limiting conditions respectively. This simple analysis illustrates that the impact of interfacial spin relaxation might have a different impact on the FM case.}



\begin{thebibliography}{43}
\expandafter\ifx\csname natexlab\endcsname\relax\def\natexlab#1{#1}\fi
\expandafter\ifx\csname bibnamefont\endcsname\relax
  \def\bibnamefont#1{#1}\fi
\expandafter\ifx\csname bibfnamefont\endcsname\relax
  \def\bibfnamefont#1{#1}\fi
\expandafter\ifx\csname citenamefont\endcsname\relax
  \def\citenamefont#1{#1}\fi
\expandafter\ifx\csname url\endcsname\relax
  \def\url#1{\texttt{#1}}\fi
\expandafter\ifx\csname urlprefix\endcsname\relax\def\urlprefix{URL }\fi
\providecommand{\bibinfo}[2]{#2}
\providecommand{\eprint}[2][]{\url{#2}}

\bibitem[{\citenamefont{Wolf et~al.}(2001)\citenamefont{Wolf, Awschalom,
  Buhrman, Daughton, von Molnar, Roukes, Chtchelkanova, and Treger}}]{wolf01}
\bibinfo{author}{\bibfnamefont{S.~A.} \bibnamefont{Wolf}},
  \bibinfo{author}{\bibfnamefont{D.~D.} \bibnamefont{Awschalom}},
  \bibinfo{author}{\bibfnamefont{R.~A.} \bibnamefont{Buhrman}},
  \bibinfo{author}{\bibfnamefont{J.~M.} \bibnamefont{Daughton}},
  \bibinfo{author}{\bibfnamefont{S.}~\bibnamefont{von Molnar}},
  \bibinfo{author}{\bibfnamefont{M.~L.} \bibnamefont{Roukes}},
  \bibinfo{author}{\bibfnamefont{A.~Y.} \bibnamefont{Chtchelkanova}},
  \bibnamefont{and} \bibinfo{author}{\bibfnamefont{D.~M.}
  \bibnamefont{Treger}}, \bibinfo{journal}{Science}
  \textbf{\bibinfo{volume}{294}}, \bibinfo{pages}{1488} (\bibinfo{year}{2001}).

\bibitem[{\citenamefont{\v{Z}uti\'{c} et~al.}(2004)\citenamefont{\v{Z}uti\'{c},
  Fabian, and Sarma}}]{zutic04}
\bibinfo{author}{\bibfnamefont{I.}~\bibnamefont{\v{Z}uti\'{c}}},
  \bibinfo{author}{\bibfnamefont{J.}~\bibnamefont{Fabian}}, \bibnamefont{and}
  \bibinfo{author}{\bibfnamefont{S.~D.} \bibnamefont{Sarma}},
  \bibinfo{journal}{Rev. of Mod. Phys.} \textbf{\bibinfo{volume}{76}},
  \bibinfo{pages}{323} (\bibinfo{year}{2004}).

\bibitem[{itr(2007)}]{itrs07}
\bibinfo{journal}{Semiconductor Industry Association, International Technology
  Roadmap for Semiconductors, http://public.itrs.net/}  (\bibinfo{year}{2007}).

\bibitem[{\citenamefont{Nikonov et~al.}(2008)\citenamefont{Nikonov, Bourianoff,
  and Gargini}}]{nikonov08}
\bibinfo{author}{\bibfnamefont{D.~E.} \bibnamefont{Nikonov}},
  \bibinfo{author}{\bibfnamefont{G.~I.} \bibnamefont{Bourianoff}},
  \bibnamefont{and} \bibinfo{author}{\bibfnamefont{P.~A.}
  \bibnamefont{Gargini}}, \bibinfo{journal}{J. of Nanoelectronics and
  Optoelectronics} \textbf{\bibinfo{volume}{3}}, \bibinfo{pages}{3}
  (\bibinfo{year}{2008}).

\bibitem[{\citenamefont{Datta and Das}(1990)}]{datta90}
\bibinfo{author}{\bibfnamefont{S.}~\bibnamefont{Datta}} \bibnamefont{and}
  \bibinfo{author}{\bibfnamefont{B.}~\bibnamefont{Das}},
  \bibinfo{journal}{Appl. Phys. Lett.} \textbf{\bibinfo{volume}{56}},
  \bibinfo{pages}{665} (\bibinfo{year}{1990}).

\bibitem[{\citenamefont{Fert and Bruno}(1994)}]{fert94}
\bibinfo{author}{\bibfnamefont{A.}~\bibnamefont{Fert}} \bibnamefont{and}
  \bibinfo{author}{\bibfnamefont{P.}~\bibnamefont{Bruno}},
  \bibinfo{journal}{Ultrathin Magnetic Structures II, Springer}
  p.~\bibinfo{pages}{82} (\bibinfo{year}{1994}).

\bibitem[{\citenamefont{Bychkov and Rashba}(1984)}]{bychkov84}
\bibinfo{author}{\bibfnamefont{Y.~A.} \bibnamefont{Bychkov}} \bibnamefont{and}
  \bibinfo{author}{\bibfnamefont{E.~I.} \bibnamefont{Rashba}},
  \bibinfo{journal}{J. of Phys. C} \textbf{\bibinfo{volume}{17}},
  \bibinfo{pages}{6039} (\bibinfo{year}{1984}).

\bibitem[{\citenamefont{Meier et~al.}(2007)\citenamefont{Meier, Salis,
  Shorubalko, Gini, Schon, and Ensslin}}]{meier07}
\bibinfo{author}{\bibfnamefont{L.}~\bibnamefont{Meier}},
  \bibinfo{author}{\bibfnamefont{G.}~\bibnamefont{Salis}},
  \bibinfo{author}{\bibfnamefont{I.}~\bibnamefont{Shorubalko}},
  \bibinfo{author}{\bibfnamefont{E.}~\bibnamefont{Gini}},
  \bibinfo{author}{\bibfnamefont{S.}~\bibnamefont{Schon}}, \bibnamefont{and}
  \bibinfo{author}{\bibfnamefont{K.}~\bibnamefont{Ensslin}},
  \bibinfo{journal}{Nature Phys.} \textbf{\bibinfo{volume}{3}},
  \bibinfo{pages}{650} (\bibinfo{year}{2007}).

\bibitem[{\citenamefont{Adelmann et~al.}(2005)\citenamefont{Adelmann, Lou,
  Strand, Palmstrom, and Crowell}}]{adelmann05}
\bibinfo{author}{\bibfnamefont{C.}~\bibnamefont{Adelmann}},
  \bibinfo{author}{\bibfnamefont{X.}~\bibnamefont{Lou}},
  \bibinfo{author}{\bibfnamefont{J.}~\bibnamefont{Strand}},
  \bibinfo{author}{\bibfnamefont{C.~J.} \bibnamefont{Palmstrom}},
  \bibnamefont{and} \bibinfo{author}{\bibfnamefont{P.~A.}
  \bibnamefont{Crowell}}, \bibinfo{journal}{Phys. Rev. B}
  \textbf{\bibinfo{volume}{71}}, \bibinfo{pages}{121301}
  (\bibinfo{year}{2005}).

\bibitem[{\citenamefont{Jonker et~al.}(2007)\citenamefont{Jonker, Kioseoglou,
  Hanbicki, Li, and Thompson}}]{jonker07}
\bibinfo{author}{\bibfnamefont{B.~T.} \bibnamefont{Jonker}},
  \bibinfo{author}{\bibfnamefont{G.}~\bibnamefont{Kioseoglou}},
  \bibinfo{author}{\bibfnamefont{A.~T.} \bibnamefont{Hanbicki}},
  \bibinfo{author}{\bibfnamefont{C.~H.} \bibnamefont{Li}}, \bibnamefont{and}
  \bibinfo{author}{\bibfnamefont{P.~E.} \bibnamefont{Thompson}},
  \bibinfo{journal}{Nature Phys.} \textbf{\bibinfo{volume}{3}},
  \bibinfo{pages}{542} (\bibinfo{year}{2007}).

\bibitem[{\citenamefont{Huang et~al.}(2007)\citenamefont{Huang, Monsma, and
  Appelbaum}}]{huang07}
\bibinfo{author}{\bibfnamefont{B.}~\bibnamefont{Huang}},
  \bibinfo{author}{\bibfnamefont{D.~J.} \bibnamefont{Monsma}},
  \bibnamefont{and}
  \bibinfo{author}{\bibfnamefont{I.}~\bibnamefont{Appelbaum}},
  \bibinfo{journal}{Appl. Phys. Lett.} \textbf{\bibinfo{volume}{91}},
  \bibinfo{pages}{072501} (\bibinfo{year}{2007}).

\bibitem[{\citenamefont{Appelbaum et~al.}(2007)\citenamefont{Appelbaum, Huang,
  and Monsma}}]{appelbaum07}
\bibinfo{author}{\bibfnamefont{I.}~\bibnamefont{Appelbaum}},
  \bibinfo{author}{\bibfnamefont{B.}~\bibnamefont{Huang}}, \bibnamefont{and}
  \bibinfo{author}{\bibfnamefont{D.~J.} \bibnamefont{Monsma}},
  \bibinfo{journal}{Nature Lett.} \textbf{\bibinfo{volume}{447}},
  \bibinfo{pages}{295} (\bibinfo{year}{2007}).

\bibitem[{\citenamefont{Sugahara and Tanaka}(2004)}]{sugahara04}
\bibinfo{author}{\bibfnamefont{S.}~\bibnamefont{Sugahara}} \bibnamefont{and}
  \bibinfo{author}{\bibfnamefont{M.}~\bibnamefont{Tanaka}},
  \bibinfo{journal}{Appl. Phys. Lett.} \textbf{\bibinfo{volume}{84}},
  \bibinfo{pages}{2307} (\bibinfo{year}{2004}).

\bibitem[{\citenamefont{de~Groot et~al.}(1983)\citenamefont{de~Groot, Mueller,
  van Engen, and Buschow}}]{groot83}
\bibinfo{author}{\bibfnamefont{R.~A.} \bibnamefont{de~Groot}},
  \bibinfo{author}{\bibfnamefont{F.~M.} \bibnamefont{Mueller}},
  \bibinfo{author}{\bibfnamefont{P.~G.} \bibnamefont{van Engen}},
  \bibnamefont{and} \bibinfo{author}{\bibfnamefont{K.~H.~J.}
  \bibnamefont{Buschow}}, \bibinfo{journal}{Phys. Rev. Lett.}
  \textbf{\bibinfo{volume}{50}}, \bibinfo{pages}{2024} (\bibinfo{year}{1983}).

\bibitem[{\citenamefont{Tyryshkin et~al.}(2005)\citenamefont{Tyryshkin, Lyon,
  Jantsch, and Schaffler}}]{tyryshkin05}
\bibinfo{author}{\bibfnamefont{A.~M.} \bibnamefont{Tyryshkin}},
  \bibinfo{author}{\bibfnamefont{S.~A.} \bibnamefont{Lyon}},
  \bibinfo{author}{\bibfnamefont{W.}~\bibnamefont{Jantsch}}, \bibnamefont{and}
  \bibinfo{author}{\bibfnamefont{F.}~\bibnamefont{Schaffler}},
  \bibinfo{journal}{Phys. Rev. Lett.} \textbf{\bibinfo{volume}{94}},
  \bibinfo{pages}{126802} (\bibinfo{year}{2005}).

\bibitem[{\citenamefont{Tanaka and Sugahara}(2007)}]{tanaka07}
\bibinfo{author}{\bibfnamefont{M.}~\bibnamefont{Tanaka}} \bibnamefont{and}
  \bibinfo{author}{\bibfnamefont{S.}~\bibnamefont{Sugahara}},
  \bibinfo{journal}{IEEE Trans. on Elec. Dev.} \textbf{\bibinfo{volume}{54}},
  \bibinfo{pages}{961} (\bibinfo{year}{2007}).

\bibitem[{\citenamefont{Hanssen et~al.}(1990)\citenamefont{Hanssen, Mijnarends,
  Rabou, and Buschow}}]{hanssen90}
\bibinfo{author}{\bibfnamefont{K.~E. H.~M.} \bibnamefont{Hanssen}},
  \bibinfo{author}{\bibfnamefont{P.~E.} \bibnamefont{Mijnarends}},
  \bibinfo{author}{\bibfnamefont{L.~P. L.~M.} \bibnamefont{Rabou}},
  \bibnamefont{and} \bibinfo{author}{\bibfnamefont{K.~H.~J.}
  \bibnamefont{Buschow}}, \bibinfo{journal}{Phys. Rev. B}
  \textbf{\bibinfo{volume}{42}}, \bibinfo{pages}{1533} (\bibinfo{year}{1990}).

\bibitem[{\citenamefont{Mancoff et~al.}(1999)\citenamefont{Mancoff, Clemens,
  Singley, and Basov}}]{mancoff99}
\bibinfo{author}{\bibfnamefont{F.~B.} \bibnamefont{Mancoff}},
  \bibinfo{author}{\bibfnamefont{B.~M.} \bibnamefont{Clemens}},
  \bibinfo{author}{\bibfnamefont{E.~J.} \bibnamefont{Singley}},
  \bibnamefont{and} \bibinfo{author}{\bibfnamefont{D.~N.} \bibnamefont{Basov}},
  \bibinfo{journal}{Phys. Rev. B} \textbf{\bibinfo{volume}{60}},
  \bibinfo{pages}{R12565} (\bibinfo{year}{1999}).

\bibitem[{\citenamefont{Galanakis and Dederichs}(2002)}]{galanakis02}
\bibinfo{author}{\bibfnamefont{I.}~\bibnamefont{Galanakis}} \bibnamefont{and}
  \bibinfo{author}{\bibfnamefont{P.~H.} \bibnamefont{Dederichs}},
  \bibinfo{journal}{Phys. Rev. B} \textbf{\bibinfo{volume}{66}},
  \bibinfo{pages}{134428} (\bibinfo{year}{2002}).

\bibitem[{\citenamefont{de~Wijs and de~Groot}(2001)}]{wijs01}
\bibinfo{author}{\bibfnamefont{G.~A.} \bibnamefont{de~Wijs}} \bibnamefont{and}
  \bibinfo{author}{\bibfnamefont{R.~A.} \bibnamefont{de~Groot}},
  \bibinfo{journal}{Phys. Rev. B} \textbf{\bibinfo{volume}{64}},
  \bibinfo{pages}{020402} (\bibinfo{year}{2001}).

\bibitem[{\citenamefont{Attema et~al.}(2006)\citenamefont{Attema, de~Wijs, and
  de~Groot}}]{attema06}
\bibinfo{author}{\bibfnamefont{J.~J.} \bibnamefont{Attema}},
  \bibinfo{author}{\bibfnamefont{G.~A.} \bibnamefont{de~Wijs}},
  \bibnamefont{and} \bibinfo{author}{\bibfnamefont{R.~A.}
  \bibnamefont{de~Groot}}, \bibinfo{journal}{J. of Phys. D: Appl. Phys.}
  \textbf{\bibinfo{volume}{39}}, \bibinfo{pages}{793} (\bibinfo{year}{2006}).

\bibitem[{\citenamefont{Bona et~al.}(1985)\citenamefont{Bona, Meier, Taborelli,
  Bucher, and Schmidt}}]{bona85}
\bibinfo{author}{\bibfnamefont{G.~L.} \bibnamefont{Bona}},
  \bibinfo{author}{\bibfnamefont{F.}~\bibnamefont{Meier}},
  \bibinfo{author}{\bibfnamefont{M.}~\bibnamefont{Taborelli}},
  \bibinfo{author}{\bibfnamefont{E.}~\bibnamefont{Bucher}}, \bibnamefont{and}
  \bibinfo{author}{\bibfnamefont{P.~H.} \bibnamefont{Schmidt}},
  \bibinfo{journal}{Sol. State Comm.} \textbf{\bibinfo{volume}{56}},
  \bibinfo{pages}{391} (\bibinfo{year}{1985}).

\bibitem[{\citenamefont{Tanaka et~al.}(1999)\citenamefont{Tanaka, Nowak, and
  Moodera}}]{tanaka99}
\bibinfo{author}{\bibfnamefont{C.~T.} \bibnamefont{Tanaka}},
  \bibinfo{author}{\bibfnamefont{J.}~\bibnamefont{Nowak}}, \bibnamefont{and}
  \bibinfo{author}{\bibfnamefont{J.~S.} \bibnamefont{Moodera}},
  \bibinfo{journal}{J. of Appl. Phys.} \textbf{\bibinfo{volume}{86}},
  \bibinfo{pages}{6239} (\bibinfo{year}{1999}).

\bibitem[{\citenamefont{\v{Z}uti\'{c} et~al.}(2006)\citenamefont{\v{Z}uti\'{c},
  Fabian, and Erwin}}]{zutic06}
\bibinfo{author}{\bibfnamefont{I.}~\bibnamefont{\v{Z}uti\'{c}}},
  \bibinfo{author}{\bibfnamefont{J.}~\bibnamefont{Fabian}}, \bibnamefont{and}
  \bibinfo{author}{\bibfnamefont{S.~C.} \bibnamefont{Erwin}},
  \bibinfo{journal}{IBM J. Research \& Development}
  \textbf{\bibinfo{volume}{50}}, \bibinfo{pages}{121} (\bibinfo{year}{2006}).

\bibitem[{\citenamefont{Bauer et~al.}(2001)\citenamefont{Bauer, Nazarov, and
  Brataas}}]{bauer01}
\bibinfo{author}{\bibfnamefont{G.~E.~W.} \bibnamefont{Bauer}},
  \bibinfo{author}{\bibfnamefont{Y.~V.} \bibnamefont{Nazarov}},
  \bibnamefont{and} \bibinfo{author}{\bibfnamefont{A.}~\bibnamefont{Brataas}},
  \bibinfo{journal}{Physica E} \textbf{\bibinfo{volume}{11}},
  \bibinfo{pages}{137} (\bibinfo{year}{2001}).

\bibitem[{\citenamefont{Yanik et~al.}(2007)\citenamefont{Yanik, Klimeck, and
  Datta}}]{yanik07}
\bibinfo{author}{\bibfnamefont{A.~A.} \bibnamefont{Yanik}},
  \bibinfo{author}{\bibfnamefont{G.}~\bibnamefont{Klimeck}}, \bibnamefont{and}
  \bibinfo{author}{\bibfnamefont{S.}~\bibnamefont{Datta}},
  \bibinfo{journal}{Phys. Rev. B} \textbf{\bibinfo{volume}{76}},
  \bibinfo{pages}{045213} (\bibinfo{year}{2007}).

\bibitem[{\citenamefont{Salahuddin et~al.}(2007)\citenamefont{Salahuddin,
  Datta, Srivastava, and Datta}}]{salahuddin07}
\bibinfo{author}{\bibfnamefont{S.}~\bibnamefont{Salahuddin}},
  \bibinfo{author}{\bibfnamefont{D.}~\bibnamefont{Datta}},
  \bibinfo{author}{\bibfnamefont{P.}~\bibnamefont{Srivastava}},
  \bibnamefont{and} \bibinfo{author}{\bibfnamefont{S.}~\bibnamefont{Datta}},
  \bibinfo{journal}{IEEE Int. Elec. Dev. Meet.} pp. \bibinfo{pages}{121--124}
  (\bibinfo{year}{2007}).

\bibitem[{\citenamefont{Wang et~al.}(2003)\citenamefont{Wang, Wang, and
  Guo}}]{wang03}
\bibinfo{author}{\bibfnamefont{B.}~\bibnamefont{Wang}},
  \bibinfo{author}{\bibfnamefont{J.}~\bibnamefont{Wang}}, \bibnamefont{and}
  \bibinfo{author}{\bibfnamefont{H.}~\bibnamefont{Guo}},
  \bibinfo{journal}{Phys. Rev. B} \textbf{\bibinfo{volume}{67}},
  \bibinfo{pages}{092408} (\bibinfo{year}{2003}).

\bibitem[{\citenamefont{Rocha et~al.}(2005)\citenamefont{Rocha, Garcia-Suarez,
  Bailey, Lambert, Ferrer, and Sanvito}}]{rocha05}
\bibinfo{author}{\bibfnamefont{A.}~\bibnamefont{Rocha}},
  \bibinfo{author}{\bibfnamefont{V.~M.} \bibnamefont{Garcia-Suarez}},
  \bibinfo{author}{\bibfnamefont{S.~W.} \bibnamefont{Bailey}},
  \bibinfo{author}{\bibfnamefont{C.~J.} \bibnamefont{Lambert}},
  \bibinfo{author}{\bibfnamefont{J.}~\bibnamefont{Ferrer}}, \bibnamefont{and}
  \bibinfo{author}{\bibfnamefont{S.}~\bibnamefont{Sanvito}},
  \bibinfo{journal}{Nature Mat.} \textbf{\bibinfo{volume}{4}},
  \bibinfo{pages}{335} (\bibinfo{year}{2005}).

\bibitem[{\citenamefont{Datta}(1997)}]{datta97}
\bibinfo{author}{\bibfnamefont{S.}~\bibnamefont{Datta}},
  \bibinfo{journal}{Electronic Transport in Mesoscopic Systems, Cambridge
  University Press}  (\bibinfo{year}{1997}).

\bibitem[{\citenamefont{Haug and Jauho}(1996)}]{haug96}
\bibinfo{author}{\bibfnamefont{H.}~\bibnamefont{Haug}} \bibnamefont{and}
  \bibinfo{author}{\bibfnamefont{A.~P.} \bibnamefont{Jauho}},
  \bibinfo{journal}{Springer Series in Solid State Sciences 123}
  (\bibinfo{year}{1996}).

\bibitem[{\citenamefont{Mahan}(1990)}]{mahan90}
\bibinfo{author}{\bibfnamefont{G.~D.} \bibnamefont{Mahan}},
  \bibinfo{journal}{Plenum Press, New York and London}  (\bibinfo{year}{1990}).

\bibitem[{\citenamefont{Datta}(2004)}]{datta04}
\bibinfo{author}{\bibfnamefont{S.}~\bibnamefont{Datta}},
  \bibinfo{journal}{Proc. of Inter. School of Phys. Societa Italiana di Fisica}
  p. \bibinfo{pages}{244} (\bibinfo{year}{2004}).

\bibitem[{\citenamefont{Halen and Pulfrey}(1985)}]{halen85}
\bibinfo{author}{\bibfnamefont{P.~V.} \bibnamefont{Halen}} \bibnamefont{and}
  \bibinfo{author}{\bibfnamefont{D.~L.} \bibnamefont{Pulfrey}},
  \bibinfo{journal}{J. of Appl. Phys.} \textbf{\bibinfo{volume}{57}},
  \bibinfo{pages}{5271} (\bibinfo{year}{1985}).

\bibitem[{\citenamefont{Bagrets et~al.}(2002)\citenamefont{Bagrets, Bagrets,
  Vedyayev, and Dieny}}]{bagrets02}
\bibinfo{author}{\bibfnamefont{D.}~\bibnamefont{Bagrets}},
  \bibinfo{author}{\bibfnamefont{A.}~\bibnamefont{Bagrets}},
  \bibinfo{author}{\bibfnamefont{A.}~\bibnamefont{Vedyayev}}, \bibnamefont{and}
  \bibinfo{author}{\bibfnamefont{B.}~\bibnamefont{Dieny}},
  \bibinfo{journal}{Phys. Rev. B} \textbf{\bibinfo{volume}{65}},
  \bibinfo{pages}{064430} (\bibinfo{year}{2002}).

\bibitem[{\citenamefont{Gundlach}(1966)}]{gundlach66}
\bibinfo{author}{\bibfnamefont{K.~H.} \bibnamefont{Gundlach}},
  \bibinfo{journal}{Sol. State Elec.} \textbf{\bibinfo{volume}{9}},
  \bibinfo{pages}{949} (\bibinfo{year}{1966}).

\bibitem[{\citenamefont{Hammar et~al.}(1999)\citenamefont{Hammar, Bennett,
  Yang, and Johnson}}]{hammar99}
\bibinfo{author}{\bibfnamefont{P.~R.} \bibnamefont{Hammar}},
  \bibinfo{author}{\bibfnamefont{B.~R.} \bibnamefont{Bennett}},
  \bibinfo{author}{\bibfnamefont{M.~J.} \bibnamefont{Yang}}, \bibnamefont{and}
  \bibinfo{author}{\bibfnamefont{M.}~\bibnamefont{Johnson}},
  \bibinfo{journal}{Phys. Rev. Lett.} \textbf{\bibinfo{volume}{83}},
  \bibinfo{pages}{203} (\bibinfo{year}{1999}).

\bibitem[{\citenamefont{Monzon et~al.}(2000)\citenamefont{Monzon, Tang, and
  Roukes}}]{monzon00}
\bibinfo{author}{\bibfnamefont{F.~G.} \bibnamefont{Monzon}},
  \bibinfo{author}{\bibfnamefont{H.~X.} \bibnamefont{Tang}}, \bibnamefont{and}
  \bibinfo{author}{\bibfnamefont{M.~L.} \bibnamefont{Roukes}},
  \bibinfo{journal}{Phys. Rev. Lett.} \textbf{\bibinfo{volume}{84}},
  \bibinfo{pages}{5022} (\bibinfo{year}{2000}).

\bibitem[{\citenamefont{Wees}(2000)}]{wees00}
\bibinfo{author}{\bibfnamefont{B.~J.} \bibnamefont{Wees}},
  \bibinfo{journal}{Phys. Rev. Lett.} \textbf{\bibinfo{volume}{84}},
  \bibinfo{pages}{5023} (\bibinfo{year}{2000}).

\bibitem[{\citenamefont{Lou et~al.}(2007)\citenamefont{Lou, Adelmann, Crooker,
  Garlid, Zhang, Reddy, Flexner, Palmstrom, and Crowell}}]{lou07}
\bibinfo{author}{\bibfnamefont{X.}~\bibnamefont{Lou}},
  \bibinfo{author}{\bibfnamefont{C.}~\bibnamefont{Adelmann}},
  \bibinfo{author}{\bibfnamefont{S.~A.} \bibnamefont{Crooker}},
  \bibinfo{author}{\bibfnamefont{E.~S.} \bibnamefont{Garlid}},
  \bibinfo{author}{\bibfnamefont{J.}~\bibnamefont{Zhang}},
  \bibinfo{author}{\bibfnamefont{K.~S.~M.} \bibnamefont{Reddy}},
  \bibinfo{author}{\bibfnamefont{S.~D.} \bibnamefont{Flexner}},
  \bibinfo{author}{\bibfnamefont{C.~J.} \bibnamefont{Palmstrom}},
  \bibnamefont{and} \bibinfo{author}{\bibfnamefont{P.~A.}
  \bibnamefont{Crowell}}, \bibinfo{journal}{Nature Phys.}
  \textbf{\bibinfo{volume}{3}}, \bibinfo{pages}{197} (\bibinfo{year}{2007}).

\bibitem[{\citenamefont{Hammar et~al.}(2000)\citenamefont{Hammar, Bennett,
  Yang, and Johnson}}]{hammar00}
\bibinfo{author}{\bibfnamefont{P.~R.} \bibnamefont{Hammar}},
  \bibinfo{author}{\bibfnamefont{B.~R.} \bibnamefont{Bennett}},
  \bibinfo{author}{\bibfnamefont{M.~J.} \bibnamefont{Yang}}, \bibnamefont{and}
  \bibinfo{author}{\bibfnamefont{M.}~\bibnamefont{Johnson}},
  \bibinfo{journal}{Phys. Rev. Lett.} \textbf{\bibinfo{volume}{84}},
  \bibinfo{pages}{5024} (\bibinfo{year}{2000}).

\bibitem[{\citenamefont{MacLaren et~al.}(1997)\citenamefont{MacLaren, Zhang,
  and Butler}}]{maclaren97}
\bibinfo{author}{\bibfnamefont{J.~M.} \bibnamefont{MacLaren}},
  \bibinfo{author}{\bibfnamefont{X.~G.} \bibnamefont{Zhang}}, \bibnamefont{and}
  \bibinfo{author}{\bibfnamefont{W.~H.} \bibnamefont{Butler}},
  \bibinfo{journal}{Phys. Rev. B} \textbf{\bibinfo{volume}{56}},
  \bibinfo{pages}{11827} (\bibinfo{year}{1997}).

\bibitem[{\citenamefont{Venugopal et~al.}(2002)\citenamefont{Venugopal, Ren,
  Datta, and Lundstrom}}]{venugopal02}
\bibinfo{author}{\bibfnamefont{R.}~\bibnamefont{Venugopal}},
  \bibinfo{author}{\bibfnamefont{Z.}~\bibnamefont{Ren}},
  \bibinfo{author}{\bibfnamefont{S.}~\bibnamefont{Datta}}, \bibnamefont{and}
  \bibinfo{author}{\bibfnamefont{M.~S.} \bibnamefont{Lundstrom}},
  \bibinfo{journal}{J. of Appl. Phys.} \textbf{\bibinfo{volume}{92}},
  \bibinfo{pages}{3730} (\bibinfo{year}{2002}).

\end{thebibliography}

\end{document}